\RequirePackage[running]{lineno}

\documentclass[aps,prl,nofootinbib,twocolumn,showpacs,superscriptaddress,letterpaper,amsmath,amssymb]{revtex4}

\usepackage{graphicx}  
\usepackage{dcolumn}   
\usepackage{bm}        
\usepackage{amssymb}   
\usepackage{feynmf}
\usepackage{slashed}
\usepackage{multirow}
\def\gsim{\lower0.5ex\hbox{$\:\buildrel >\over\sim\:$}}
\def\lsim{\lower0.5ex\hbox{$\:\buildrel <\over\sim\:$}}

\newcommand{\be}{\begin{equation}}
\newcommand{\ee}{\end{equation}}
\newcommand{\bea}{\begin{eqnarray}}
\newcommand{\eea}{\end{eqnarray}}

\newcommand{\nbox}{{\,\lower0.9pt\vbox{\hrule \hbox{\vrule height 0.2 cm
\hskip 0.2 cm \vrule height 0.2 cm}\hrule}\,}}

\def\missET {{\not\!\! E_T}}

\begin{document}

\thispagestyle{empty}
\vspace*{-3.5cm}

\vspace{0.5in}

\title{Mono-everything: combined limits on dark matter production\\ at
colliders from multiple final states}

\begin{center}
\begin{abstract}
Searches for dark matter production at particle colliders are
complementary to direct-detection and indirect-detection experiments, and especially powerful
for small masses, $m_\chi<100$ GeV.  An important collider dark matter
signature is due to the production
of a pair of these invisible particles with the initial-state radiation of a
standard model particle. Currently, collider searches use
individual and nearly orthogonal final states to search for initial-state
jets, photons or massive gauge bosons.  We combine these results across
final states and across experiments to give the strongest current
collider-based limits in the context of effective field theories, and
map these  to limits on dark matter interactions with nuclei and
to dark matter self-annhiliation.
\end{abstract}
\end{center}

\author{Ning Zhou}
\affiliation{Department of Physics and Astronomy, University of California, Irvine, CA 92697}
\author{David Berge}
\affiliation{GRAPPA Institute, University of Amsterdam, Netherlands}
\author{Daniel Whiteson}
\affiliation{Department of Physics and Astronomy, University of
  California, Irvine, CA 92697}

\maketitle

\linenumbers


Though the presence of dark matter in the universe has been
well-established, little is known of its particle nature or its
non-gravitational interactions.  A vibrant experimental program
is searching for a weakly interacting massive particle (WIMP), denoted as
$\chi$, and interactions with standard model particles via some
as-yet-unknown mediator.   If the mediator is too heavy to be
resolved, the interaction can be modeled as an effective field theory
with a four-point interaction.

\begin{figure}
\includegraphics[width=0.5\linewidth]{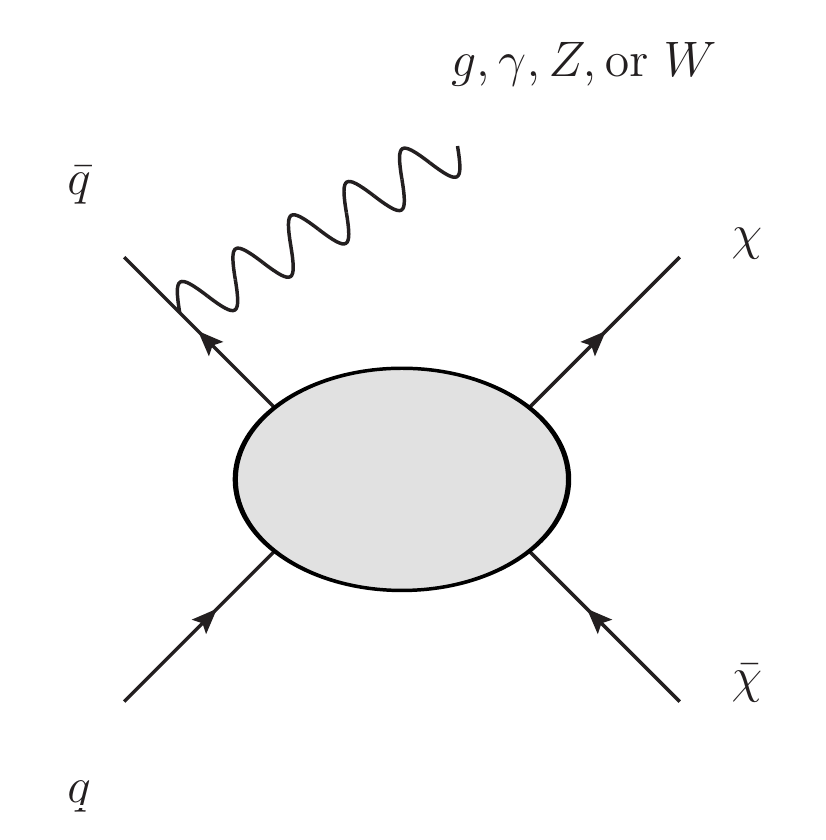}
\caption{ Pair production of WIMPs ($\chi\bar{\chi}$) in proton-proton
  collisions at the LHC via an unknown intermediate state, with initial-state radiation of a standard
  model particle.}
\label{fig:diag}
\end{figure}

One critical component of this program is the search for
pair-production of WIMPs at particle colliders, specifically
$pp\rightarrow \chi\bar{\chi}$ at the LHC via some unknown
intermediate state. As the final state WIMPs are invisible to the
detectors, the events can only be seen if there is associated
initial-state radiation of a standard model
particle~\cite{Beltran:2010ww,Fox:2011pm,Goodman:2010ku}, see Fig~\ref{fig:diag}, recoiling against the dark matter pair.

The LHC collaborations have reported limits on the cross section of
$pp\rightarrow  \chi\bar{\chi}+X$ where $X$ is a
gluon or quark~\cite{atlasjet,cmsjet}, photon~\cite{atlasphoton,cmsphoton}, and
other searches have been repurposed to study the cases where $X$ is a
$W$~\cite{monow} or $Z$ boson~\cite{atlaszz,monoz}. In each case,
limits are reported in terms of the mass scale
$M_\star$ of the unknown interaction expressed in an effective field
theory~\cite{Beltran:2008xg,Beltran:2010ww, Fox:2011pm,Goodman:2010ku, Shepherd:2009sa,Cao:2009uw,Goodman:2010yf,Bai:2010hh,Rajaraman:2011wf,Cotta:2012nj,Petriello:2008pu,Gershtein:2008bf}.
These various initial-state tags probe the same effective theory, but
are largely statistically independent due to their nearly orthogonal
event selection requirements.  As the relative rates of radiation of
gluons (quarks), photons, $W$
or $Z$ bosons from the incoming quark (gluon) legs are determined by the
standard model, the various probes may be combined to give the
strongest limits without any loss of generality or additional
theoretical assumptions.  

Recently, an analysis of multi-jet final states was shown to add some
sensitivity to the mono-jet analyses~\cite{razor}; that sample is not
statistically independent from the mono-jet results used here, and is
not included. An earlier global analysis of indirect and direct
constraints with Tevatron data and mono-jet data from ATLAS provided an initial set of combined
constraints~\cite{Cheung:2012gi} using the approximations of a
$\chi^2$ technique. 

In this paper, we perform a full statistical
combination of the limits from all available channels (mono-jet,
mono-photon, mono-$Z$\footnote{Final states with a heavy boson have
  little power relative to mono-photon or mono-jet; we include
  mono-$Z$ as a demonstration, and do not include mono-$W$, although
  see~\cite{monow}.  For an alternative view of mono-$Z$, see Ref~\cite{Bell:2012rg}}  from both ATLAS and CMS at
$\sqrt{s}=7$ TeV, accounting for the dominant correlations and
providing the most powerful current collider constraints.  While the
limits reported by the experimental collaborations are typically given for a few
select effective operators, we calculate the efficiencies of their
selections and reinterpret their searches for the complete set of
operators relevant for Dirac fermion or complex scalar WIMPs.

\subsection{Models}

\begin{figure}
\includegraphics[width=2in]{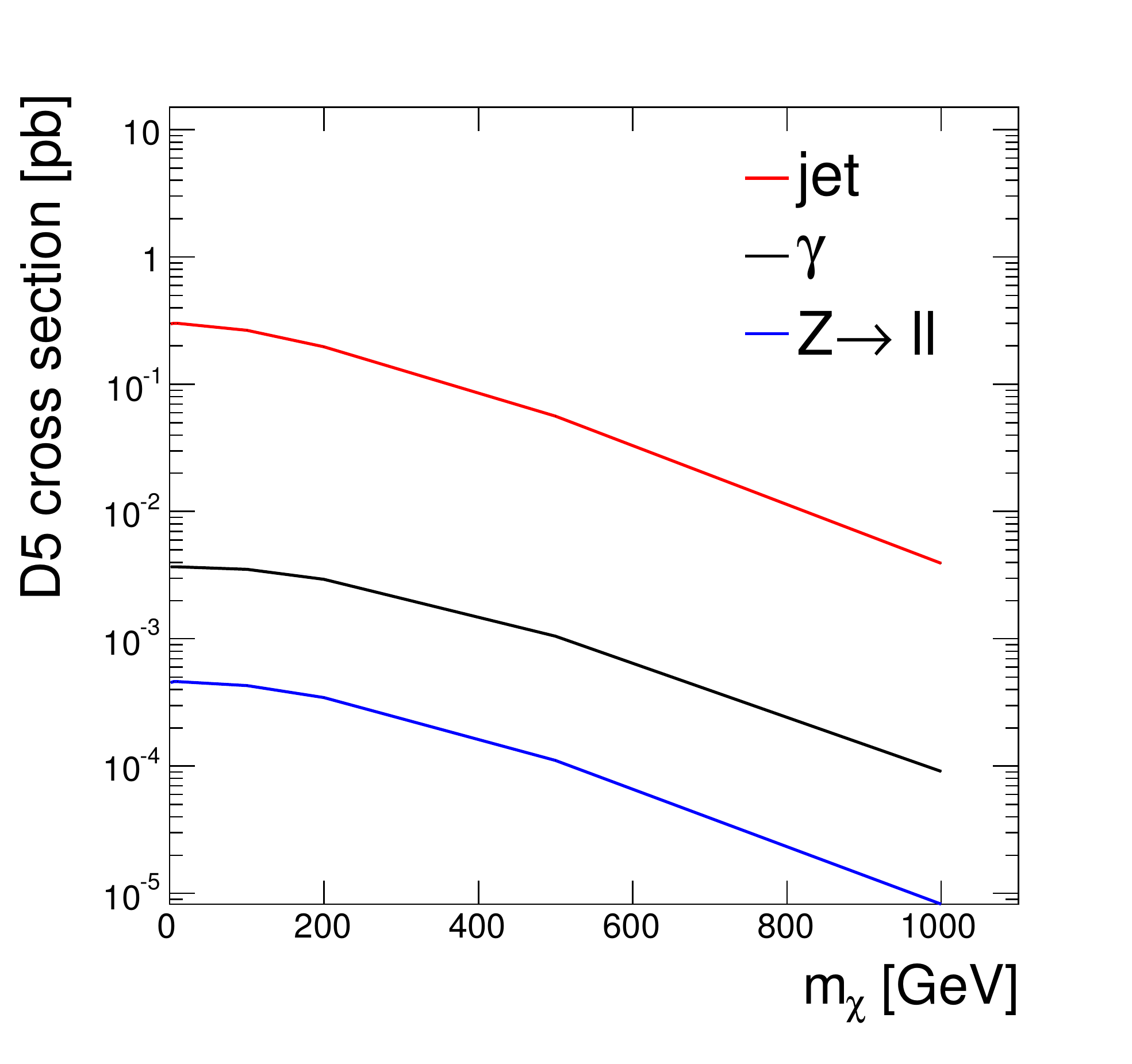}
\includegraphics[width=2in]{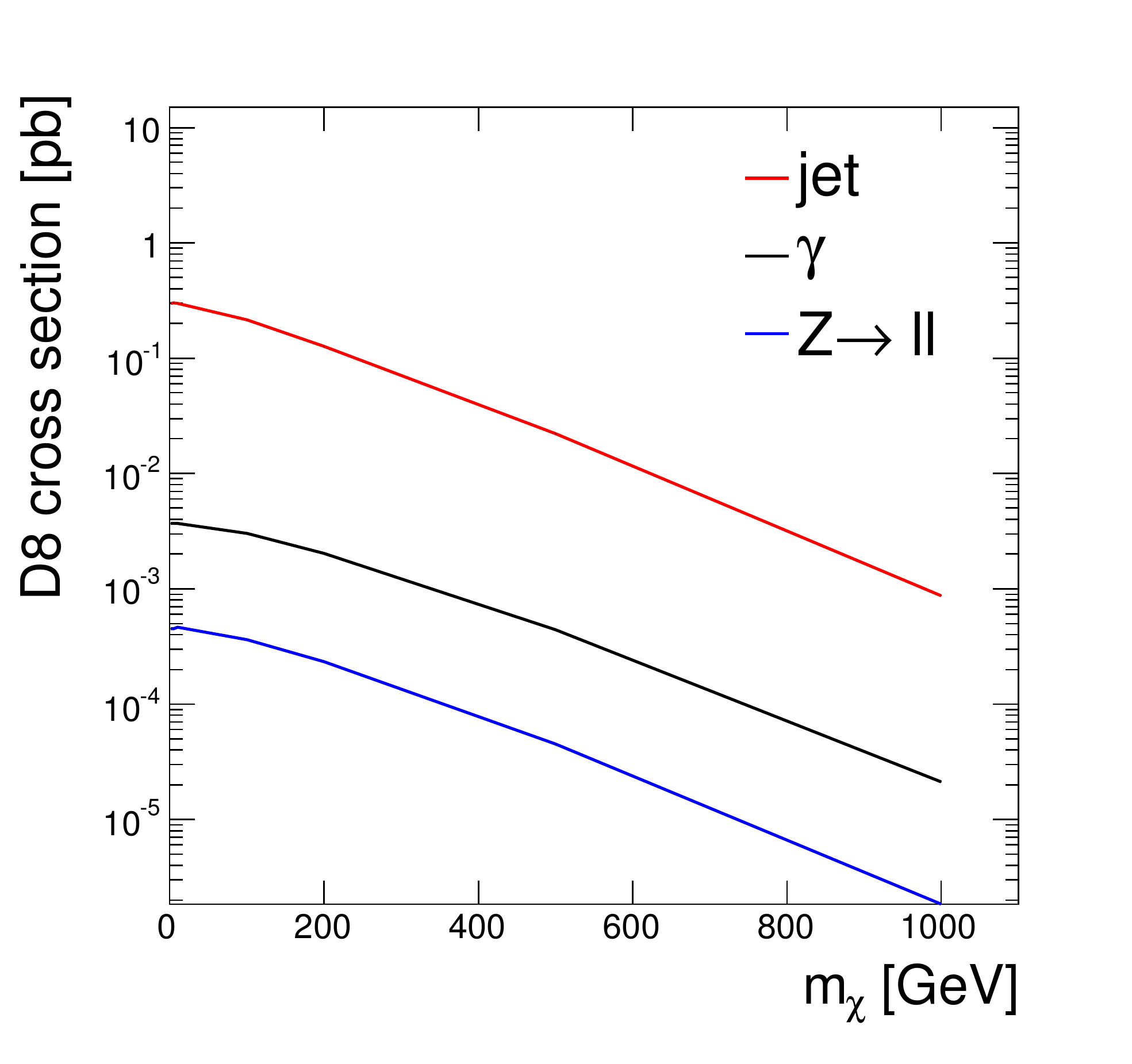}
\includegraphics[width=2in]{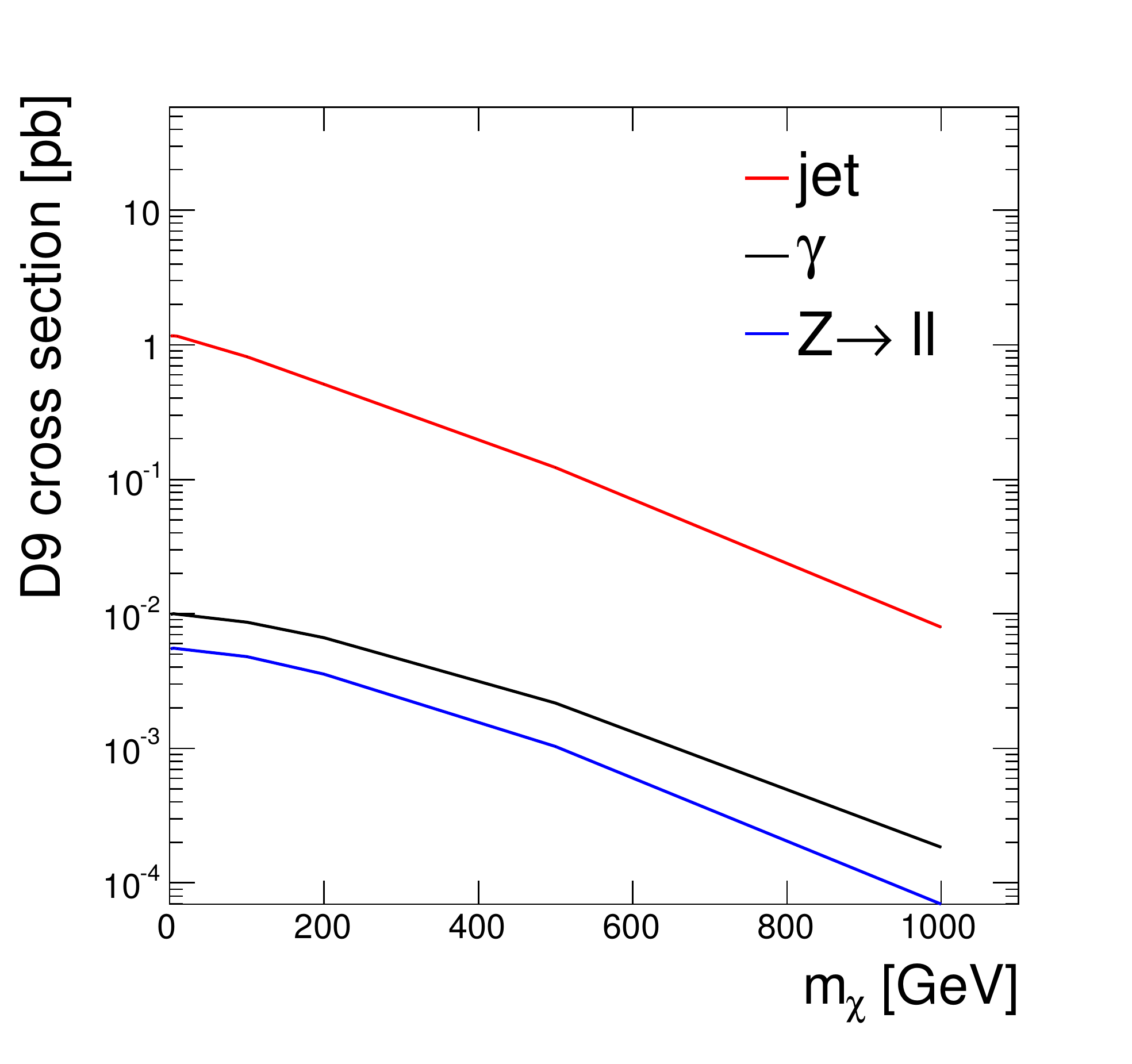}
\caption{Cross sections for $pp\rightarrow \chi\bar{\chi}+X$ production
  where $X$ is initial-state radiation of a jet, photon or $Z$
  boson. Jet and photon final states include a $p_T>80$ GeV cut at
  parton-level. Each pane shows the cross section for a different effective
  operator: top is D5, center is D8, bottom is D9. See
  Ref.~\cite{Goodman:2010ku} for operator definitions.}
\label{fig:xsec}
\end{figure}

The effective theories of dark matter considered here consider the
possibility that the final-state WIMPs are a Dirac fermion (operators D1-D14 in Ref~\cite{Goodman:2010yf}) or a
complex scalar (operators C1-C6 in Ref~\cite{Goodman:2010yf}). These
four-point effective operators assume that the unknown intermediate
particles have a heavy mass scale; we use a suppression scale,
$M_\star$.  Cross
sections at leading order for
production in $pp$ collisions at $\sqrt{s}=7$ TeV are shown in
Fig~\ref{fig:xsec} for select operators with $M_\star=1$ TeV for
illustration.  Recently, next-to-leading-order calculations have been
performed for mono-jet and mono-photon processes~\cite{Fox:2012ru}
showing ratios of $\sigma_{\textrm{NLO}}/\sigma_{\textrm{LO}} \approx
1.2-1.5$; our mono-jet results partially include this effect by
generating and matching multiple-parton emission.


For some operators, cross sections of dark matter production at the LHC
can be transformed into cross sections for WIMP-nucleon interaction,
$\sigma(\chi -n)$~\cite{Goodman:2010ku}, or WIMP
annihilations~\cite{Fox:2011pm}. Therefore the effective field
theories allow us to map measurements performed at the LHC to the
quantities relevant for direct-detection and indirect-detection dark
matter search experiments.

The effective-field-theory approach is valid as long as the unknown
new mediator particles that couple the dark-matter particles to SM
quarks or gluons are too heavy to be resolved: $q < M^*$, where $q$ is
the momentum transfer. The breakdown of the effective approach depends
ultimately on the details of the new and unknown physics, specifically
on the number of new mediator particles and the new couplings.
Therefore, these theories cannot be treated generically and must be
interpretted with some care.  To guide the interpretation, we indicate
the range of validity as lower bounds on the mass suppression scale
$M_\star$ following ref.~\cite{Goodman:2010ku}. We note that any range
of validity of the effective field theory involves assumptions about
the unknown physics, see Refs~\cite{unitary} and \cite{razor} for additional unitarity arguments and  more stringent validity ranges.

 Assuming the simplest
possible structure of new physics (mediation via exactly one new heavy
mediator of mass $M$, $M_\star = M / \sqrt{g_1\\ g_2}$, $g_1$ and $g_2$
being coupling constants), bounds on the suppression scale can be
placed by requiring $M>2m_\chi$ and that the new physics be as
strongly coupled as possible for it to be still perturbative
($\sqrt{g_1\\ g_2} < 4\pi$):
\begin{eqnarray*}
&& M_\star > \frac{m_\chi}{2\pi} (\mathrm{D5~to~D14~and~C3~to~C6}),\\
&&\sqrt{\frac{M_\star^3}{m_q}} > \frac{m_\chi}{2\pi} \,\,\, (\mathrm{D1~to~D4}),\\
&&\frac{M_\star^2}{m_q} > \frac{m_\chi}{8 \pi^2} \,\,\, (\mathrm{C1~and~C2}).
\end{eqnarray*}
Note that we are accounting for additional factors of $m_q$ in the
definitions of operators D1 to D4 and C1, C2 of
ref.~\cite{Goodman:2010ku}.

\subsection{Experimental Searches}

The experimental searches typically require one or more
high-$p_{\mathrm{T}}$ object and missing transverse momentum, see
Table~\ref{tab:cuts} for a summary and comparison of the mono-photon and mono-jet selections.

\begin{table}
\caption{Summary of event selection requirements in \mbox{ATLAS} and \mbox{CMS}
  mono-jet or mono-photon analyses. Note that ATLAS uses two signal
  regions ($\missET>350$ or $500$ GeV) for the mono-jet analyses, depending on the operator. }
  \label{tab:cuts}
\begin{tabular}{l|l|l}
\hline\hline
& ATLAS & CMS \\
\hline
jet & 1 or 2 jets& 1 or 2 jets\\
&  $p^{j_1}_{\textrm{T}}> 350 (500)$ GeV &
$p^{j_1}_{\textrm{T}}> 110$ GeV \\

&  $p^{{j_2}}_{\textrm{T}}>  30$ GeV &
$p^{{j_2}}_{\textrm{T}}> 30$ GeV \\
& $\missET>350 (500)$ GeV & $\missET>350$ GeV \\
& veto leptons& veto leptons\\
& $\Delta\phi(j_2,\missET)>0.5$ & $\Delta\phi(j_1,j_2)<2.5$ \\
\hline
$\gamma$ & 1 photon, $p_{\textrm{T}}>150$ GeV & 1 photon
$p_{\textrm{T}}>145$ GeV \\
& $\missET>150$ GeV & $\missET>130$ GeV \\
&$\le 1$ jet with $p_{\textrm{T}}>30$ GeV &$0$ track with $p_{T}>20$
GeV \\
& isolation details& isolation details\\
& $\Delta\phi(\gamma,\missET)>0.4$ & \\
& $\Delta\phi(j_1,\missET)>0.4 $ & \\
& veto leptons & \\
\hline\hline
\end{tabular}
\end{table}

The mono-$Z$ analysis~\cite{monoz} uses the ATLAS $ZZ\rightarrow
\ell\ell\nu\nu$ cross-section measurement~\cite{atlaszz}, which requires:
\begin{itemize}
\item two same-flavor opposite-sign electrons or muons, each with $p_{\rm
    T}^{\ell} > 20$ GeV, $|\eta^{\ell}|<2.5$;
\item dilepton invariant mass close to the $Z$ boson mass:
  $m_{\ell\ell} \in [m_{Z}-15,m_Z+15]$ GeV;
\item no particle-level jet with $p^j_{\rm T} >$ 25 GeV and
  $|\eta^j|<$4.5;
\item $(|p_{\rm T}^{\nu\bar{\nu}} - p_{\rm T}^Z|)/ p_{\rm T}^Z < 0.6$;
\item $-p_{\rm T}^{\nu\bar{\nu}} \times \cos( \Delta\phi(p_{\rm
    T}^{\nu\bar{\nu}},p_{\rm T}^Z) ) > 80$ GeV.
\end{itemize}

The selection efficiency of each selection for each operator is given
in Table~\ref{tab:eff} and were estimated in the
following way. References~\cite{atlasjet,cmsjet,atlasphoton,cmsphoton}
provide signal efficiency for several select operators; this
efficiency is the product of geometric and kinematic acceptance of the
selection criteria and object reconstruction efficiency. The object
reconstruction efficiency depends on the details of the detector
performance, but is largely independent of operator.  The  geometric
and kinematic acceptances  can be reliably estimated
using parton-level simulated event samples.   We measure the geometric
and kinematic efficiency for each operator, and use the quoted total efficiences to deduce the
object reconstruction efficiencies.  This allows us to estimate the
total efficiency for each operator.

\begin{table}
\caption{Selection efficiency as percentages for each channel of the analyses used in
  the combination, for operators
  D1-14 and C1-C6 for low and high values of the WIMP mass
  $m_\chi$. ATLAS mono-jet analysis has two signal regions, we use
  $\missET>500 (350)$ GeV and $p^{j_1}_{\textrm{T}}> 500 (350)$ GeV
  region for operators D9-D14 (D1-D8 and C1-C6). Operators D11-14, C5
  and C6 only couple to gluon initial states, and so have no efficiency
  for photon or $Z$ boson radiation. The $Z$ efficiencies include the
  $Z\rightarrow\ell\ell$ branching fraction. Jet and photon samples include a $p_T>80$ GeV cut at
  parton-level.}
  \label{tab:eff}
\begin{tabular}{lr|rrr|rr}
\hline
\hline
& & \multicolumn{3}{c|}{ATLAS} & \multicolumn{2}{c}{CMS}\\
Operator & $m_{\chi}$  & jet & $\gamma$ & $Z$ & jet & $\gamma$ \\
\hline
D1       & 10    & 0.4\%         & 11.2\%        & 1.2\%         & 0.7\%         & 8.0\%\\ 
         & 1000  & 2.6\%         & 19.1\%        & 1.2\%         & 3.6\%         & 11.3\%\\ \hline 
D2       & 10    & 0.4\%         & 10.8\%        & 1.2\%         & 0.7\%         & 8.0\%\\ 
         & 1000  & 2.4\%         & 18.6\%        & 1.1\%         & 3.7\%         & 11.3\%\\ \hline 
D3       & 10    & 0.5\%         & 11.1\%        & 1.2\%         & 0.7\%         & 8.0\%\\ 
         & 1000  & 2.6\%         & 18.9\%        & 1.2\%         & 3.9\%         & 11.3\%\\ \hline 
D4       & 10    & 0.5\%         & 10.8\%        & 1.2\%         & 0.7\%         & 7.6\%\\ 
         & 1000  & 2.6\%         & 18.6\%        & 1.1\%         & 3.7\%         & 11.3\%\\ \hline 
D5       & 10    & 1.7\%         & 18.2\%        & 0.9\%         & 2.2\%         & 11.3\%\\
         & 1000  & 3.3\%         & 23.5\%        & 1.1\%         & 4.5\%         & 14.7\%\\ \hline 
D6       & 10    & 1.7\%         & 18.7\%        & 0.9\%         & 2.2\%         & 12.0\%\\
         & 1000  & 3.2\%         & 23.6\%        & 1.1\%         & 4.4\%         & 15.2\%\\ \hline 
D7       & 10    & 1.7\%         & 18.1\%        & 0.9\%         & 2.4\%         & 11.3\%\\
         & 1000  & 3.3\%         & 23.4\%        & 1.1\%         & 4.4\%         & 14.5\%\\ \hline 
D8       & 10    & 1.7\%         & 18.5\%        & 0.9\%         & 2.3\%         & 11.8\%\\
         & 1000  & 3.1\%         & 23.6\%        & 1.1\%         & 4.3\%         & 15.1\%\\ \hline 
D9       & 10    & 0.9\%         & 23.5\%        & 1.4\%         & 4.1\%         & 14.1\%\\
         & 1000  & 1.2\%         & 23.3\%        & 1.4\%         & 5.1\%         & 14.8\%\\ \hline 
D10      & 10    & 1.1\%         & 23.6\%        & 1.4\%         & 4.2\%         & 14.4\%\\
         & 1000  & 1.2\%         & 23.4\%        & 1.4\%         & 5.2\%         & 14.8\%\\ \hline 
D11      & 10    & 0.9\%         & --         & --         & 4.1\%         & --\\ 
         & 1000  & 2.4\%         & --         & --         & 7.5\%         & --\\ \hline 
D12      & 10    & 1.0\%         & --         & --         & 4.2\%         & --\\ 
         & 1000  & 2.4\%         & --         & --         & 7.4\%         & --\\ \hline 
D13      & 10    & 0.9\%         & --         & --         & 4.1\%         & --\\ 
         & 1000  & 2.4\%         & --         & --         & 7.5\%         & --\\ \hline 
D14      & 10    & 1.1\%         & --         & --         & 4.0\%         & --\\ 
         & 1000  & 2.4\%         & --         & --         & 7.4\%         & --\\ \hline 
C1       & 10    & 0.1\%         & 7.0\%         & 1.0\%         & 0.2\%         & 5.3\%\\ 
         & 1000  & 2.3\%         & 18.2\%        & 1.1\%         & 3.3\%         & 11.0\%\\ \hline 
C2       & 10    & 0.1\%         & 7.0\%         & 1.0\%         & 0.1\%         & 5.6\%\\ 
         & 1000  & 2.5\%         & 18.4\%        & 1.1\%         & 3.8\%         & 11.2\%\\ \hline 
C3       & 10    & 1.7\%         & 18.4\%        & 0.9\%         & 2.3\%         & 11.6\%\\
         & 1000  & 2.9\%         & 23.6\%        & 1.1\%         & 4.1\%         & 14.9\%\\ \hline 
C4       & 10    & 1.4\%         & 18.4\%        & 0.9\%         & 2.2\%         & 11.8\%\\
         & 1000  & 3.0\%         & 23.8\%        & 1.1\%         & 4.1\%         & 15.3\%\\ \hline 
C5       & 10    & 1.4\%         & --         & --         & 1.7\%         & --\\ 
         & 1000  & 5.9\%         & --         & --         & 7.6\%\\ \hline 
C6       & 10    & 1.2\%         & --         & --         & 1.7\%         & --\\ 
         & 1000  & 5.9\%         & --         & --         & 7.6\%         & --\\ \hline 
\hline
\end{tabular}
\end{table}

%




\begin{table}
\caption{ 90\% CL limits on $N_{\rm events}$, efficiencies for
  $m_\chi=10$ GeV, and limits on $\sigma(pp\rightarrow \chi\bar{\chi}+X)$
  using the D5 operator. In the case of the $Z+\missET$ final state,
  the efficiency is relative to $Z\rightarrow\ell\ell$ decays only.}
\label{tab:exp}
\begin{tabular}{lllrrrr}
\hline\hline
Channel & Bg. & Obs & Limit & Eff. & Lumi. & Limit \\
& & & $N$ & & {\tiny(fb$^{-1}$)}& $\sigma$ (fb)\\
\hline
ATLAS jet$+\missET$ & $750\pm 60$  & 785 & 139.3 & 1.7\% & 4.8 & 1,700 \\ 
CMS jet+$\missET$ & $1225\pm 101$ & 1142 & 125.2 & 2.2\% & 5.0 & 1,140 \\
ATLAS $\gamma+\missET$ & $137\pm 20$ & 116 & 27.4 & 18\% & 4.6 & 33 \\
CMS $\gamma+\missET$ & $75.1\pm 9.4$ & 73 & 19.3 & 11\% & 5.0 & 35\\
ATLAS $Z+\missET$ & $86.2\pm 7.2$ & 87 & 21.7 & 13\% & 4.6 & 36 \\
\hline\hline
\end{tabular}
\end{table}

\begin{table}
\caption{ 90\% CL limits on $N_{\rm events}$, efficiencies for
  $m_\chi=10$ GeV, and limits on $\sigma(pp\rightarrow \chi\bar{\chi}+X)$ using the D9 operator.}
\label{tab:exp2}
\begin{tabular}{lllrrrr}
\hline\hline
Channel & Bg. & Obs & Limit & Eff. & Lumi. & Limit \\
& & & $N$ & & {\tiny(fb$^{-1}$)}& $\sigma$ (fb)\\
\hline
ATLAS jet$+\missET$ & $83\pm 14$  & 77 & 25.5 & 0.9\% & 4.8& 590 \\ 
CMS jet+$\missET$ & $1225\pm 101$ & 1142 & 125.2 & 4.1\%& 5.0 & 610 \\
\hline\hline
\end{tabular}
\end{table}
\subsection{Combination}

The separate analyses, each of which are single-bin counting
experiments, are combined into a multi-bin counting experiment. This
allows for a coherent signal rate to be tested across channels, but 
 preserves their distinct signal-to-background ratios.  

The background estimates are taken directly from the experimental
publications, see a summary in Table~\ref{tab:exp}, and are assumed to
be uncorrelated across channels, as they are typically dominated by
channel-specific or detector-specific uncertainties. For example, in some cases
the background estimates are data-driven, and the dominant
uncertainties are in the finite statistics of independent control
samples.  Inclusion of correlations up to 20\% does not qualitatively
impact the results of the combination.

The backgrounds, their uncertainties and the observed yield can be
used to calculate a 90\% CL upper limit on the number of signal
events $N$ in the sample, see Table~\ref{tab:exp} and Table~\ref{tab:exp2}, using the CLs method~\cite{cls1,cls2}.  This value is almost
completely model independent. Translating it into a limit on
the cross section for the $pp\rightarrow \chi\bar{\chi}+X$ signal requires the
effieciency of the signal in each selection, see Table~\ref{tab:exp}.
These individual limits reproduce well the results reported by the experiments.

The signal regions are nearly orthogonal, but not exactly. For
example, the mono-jet analyses do not veto events with a photon, and
the mono-photon analyses allow the presence of one jet.  From our parton-level simulated event samples, we estimated the overlaps among different channels and 
found that the overlap fraction is less than $1\%$.

The individual analyses include signal uncertainties of up to 20\% on
the cross section, mostly due to uncertainties in jet energy
calibration and levels of initial-state radiation. These uncertainties do not affect the cross-section
limits, but can be simply applied to limits on $M_\star$. In each
case, we quote the limit using the central value.

To summarize, the assumptions made in this combination are
\begin{itemize}
\item the background uncertainties are monolithic and uncorrelated, and
\item the signal selections are orthogonal
\end{itemize}

Combining channels is then straightforward, though the intermediate step of a model-independent
limit on the number of events $N$ is no longer possible, as the
limits depend on the relative distribution of signal events across
channels, which is model specific.  Instead, cross-section limits are
obtained directly.  These limits are then converted into limits on
$M_\star$, using the relationships from Ref.~\cite{Goodman:2010yf}.
The individual-channel limits, combination across experiments and the
grand combination of all channels are shown in  Table~\ref{tab:lims}
for the D5 operator and one choice of $m_\chi$.  
Clearly the mono-jet analyses are the most powerful, and the greatest gain in combination
is from combining the ATLAS and CMS mono-jet analyses, though the
addition of the mono-photon and mono-$Z$ gives a non-negligible
improvement in the combined result.

\begin{table}
\caption{ 90\% CL limits on $\sigma(pp\rightarrow \chi\bar{\chi}+X)$ for
  $m_\chi=10$ GeV, theory
  prediction for $M_\star=1$ TeV, and limits on $M_\star$ using the D5
  operator. In the case of the $Z+\missET$ final state,
  the prections include the $Z\rightarrow\ell\ell$ branching fraction.}
\label{tab:lims}
\begin{tabular}{lrrrrrr}
\hline\hline
Channel & Limit $\sigma$ & Pred. & Limit
$M_\star$ & & \\ 
 & (fb) & (fb) & (GeV) \\ \hline
ATLAS jet$+\missET$   & 1,700 & 370 & 685 & \multirow{2}{*}{{\Large\}}
  785}& \multirow{5}{*}{{\Huge\}}}\\
CMS jet+$\missET$ & 1,140 & 370 & 750 \\
ATLAS $\gamma+\missET$ & 33  & 3.7 &  580& \multirow{2}{*}{{\Large\}}
  645}& &795 \\
CMS $\gamma+\missET$ & 35 & 3.7 & 570 \\
ATLAS $Z+\missET$  & 36 & 0.5 & 340\\
\hline\hline
\end{tabular}
\end{table}

\begin{figure}[h]
\includegraphics[width=3.in]{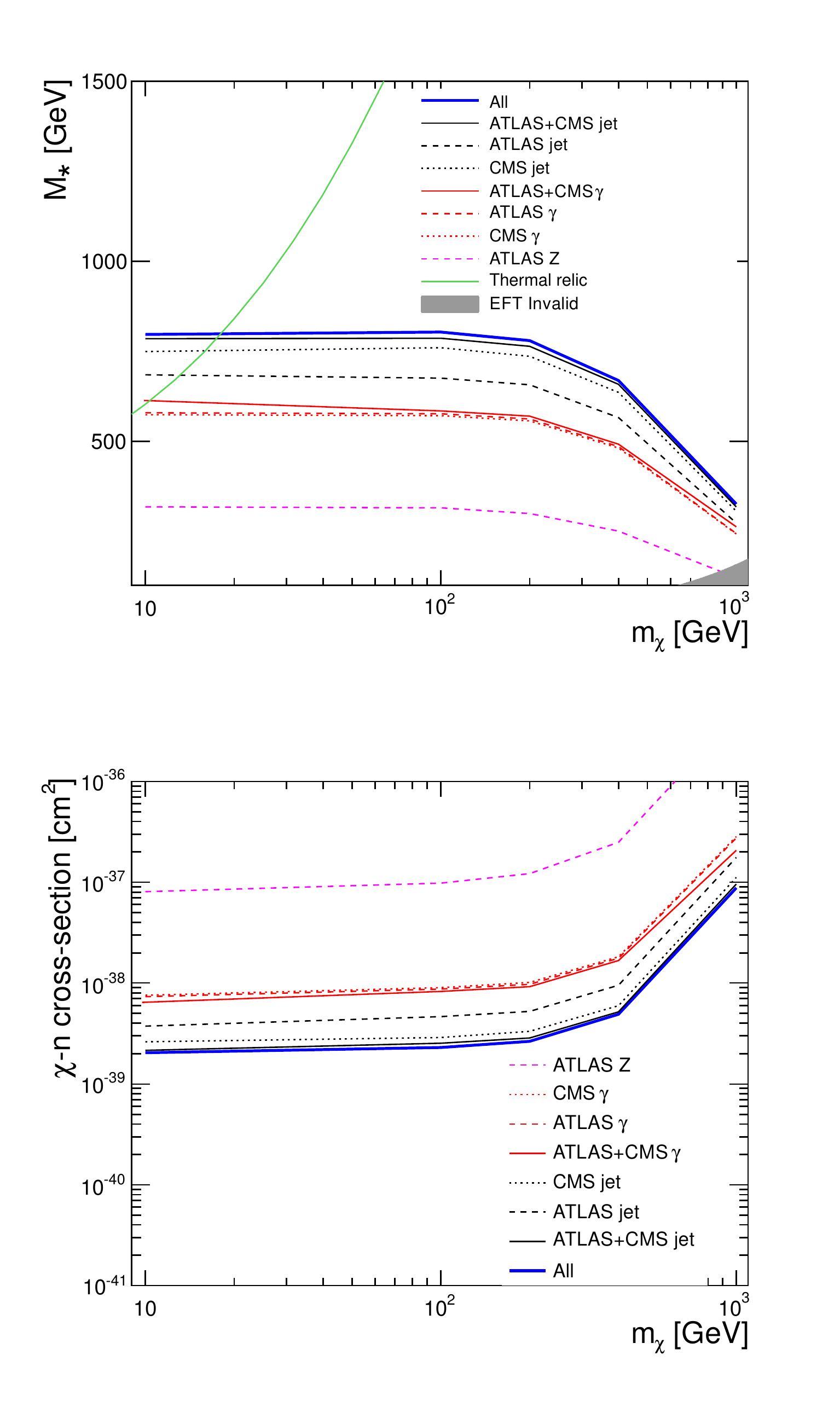}
\caption{Limits at 90\% CL in $M_\star$ (top) and in the spin-independent
  WIMP-nucleon cross section (bottom) for individual and combined limits using
  the D5 operator as a function of $m_\chi$.}
\label{fig:d5}
\end{figure}

\begin{figure}[h]
\includegraphics[width=3.in]{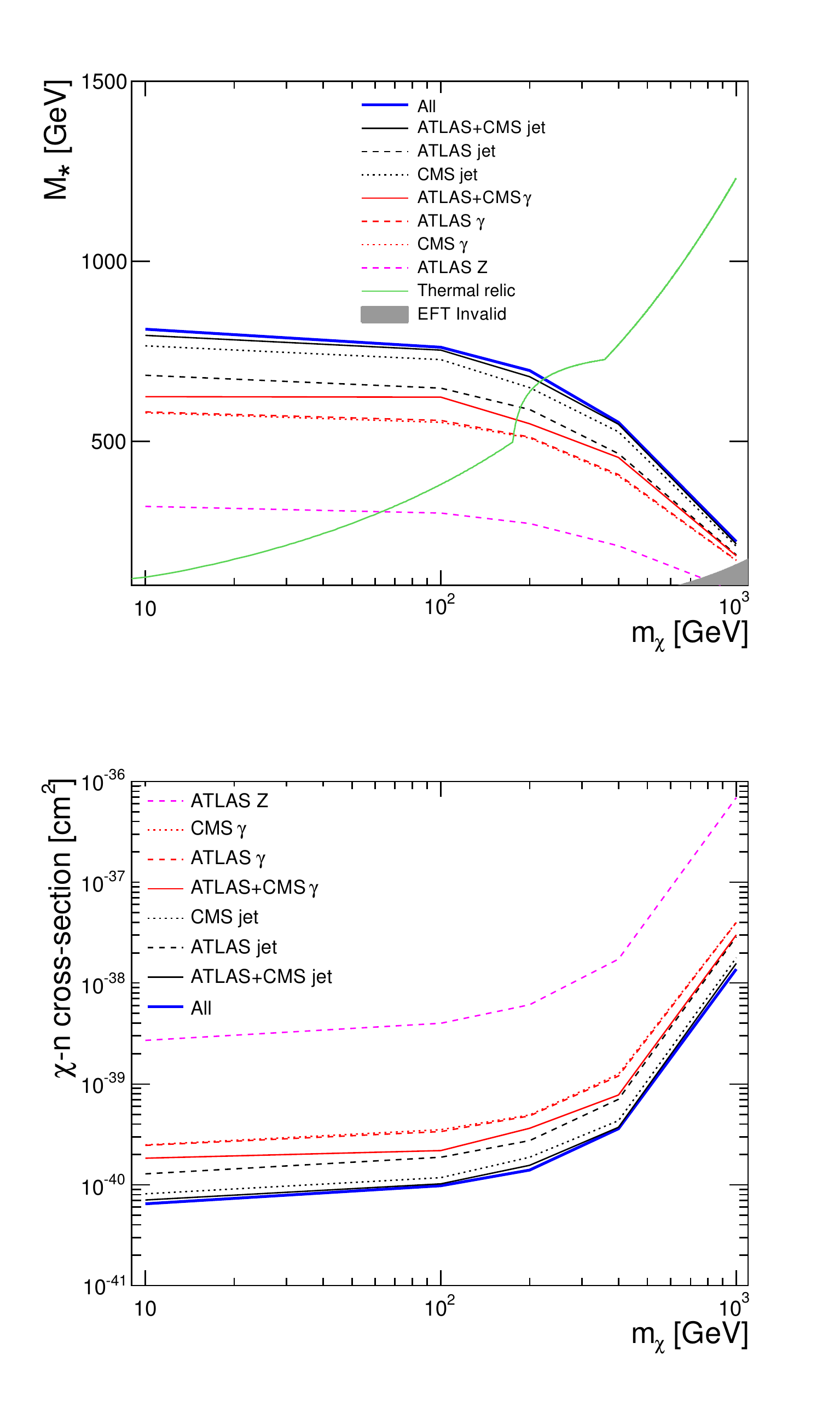}
\caption{Limits at 90\% CL in $M_\star$ (top) and in the spin-dependent
  WIMP-nucleon cross section (bottom) for individual and combined limits using
  the D8 operator as a function of $m_\chi$.}
\label{fig:d8}
\end{figure}

Limits on $M_\star$ for the D5 and D8 operators are shown in
Fig~\ref{fig:d5} and~\ref{fig:d8}, as well as limits on $\sigma(\chi
-n)$. Where the $M_\star$ limits exceed the thermal relic values taken
from Ref.~\cite{Goodman:2010ku}, assuming that dark matter is entirely
composed of thermal relics, the resulting dark matter density of the universe
would contradict WMAP measurements; therefore, WIMPs cannot couple to quarks or
gluons exclusively via the given operator and account entirely for the
relic density. This $m_\chi$ region is
either excluded, or requires that annihilation channels to leptons must
exist, or participation of different operators which interfere negatively, thereby
reducing the limits on $M_\star$.

\subsection{Application to other models}

While the experimental results are usually quoted for a small
selection of the effective operator models, the analyses are clearly
relevant for all of them. 

\begin{figure}
\includegraphics[width=2.5in]{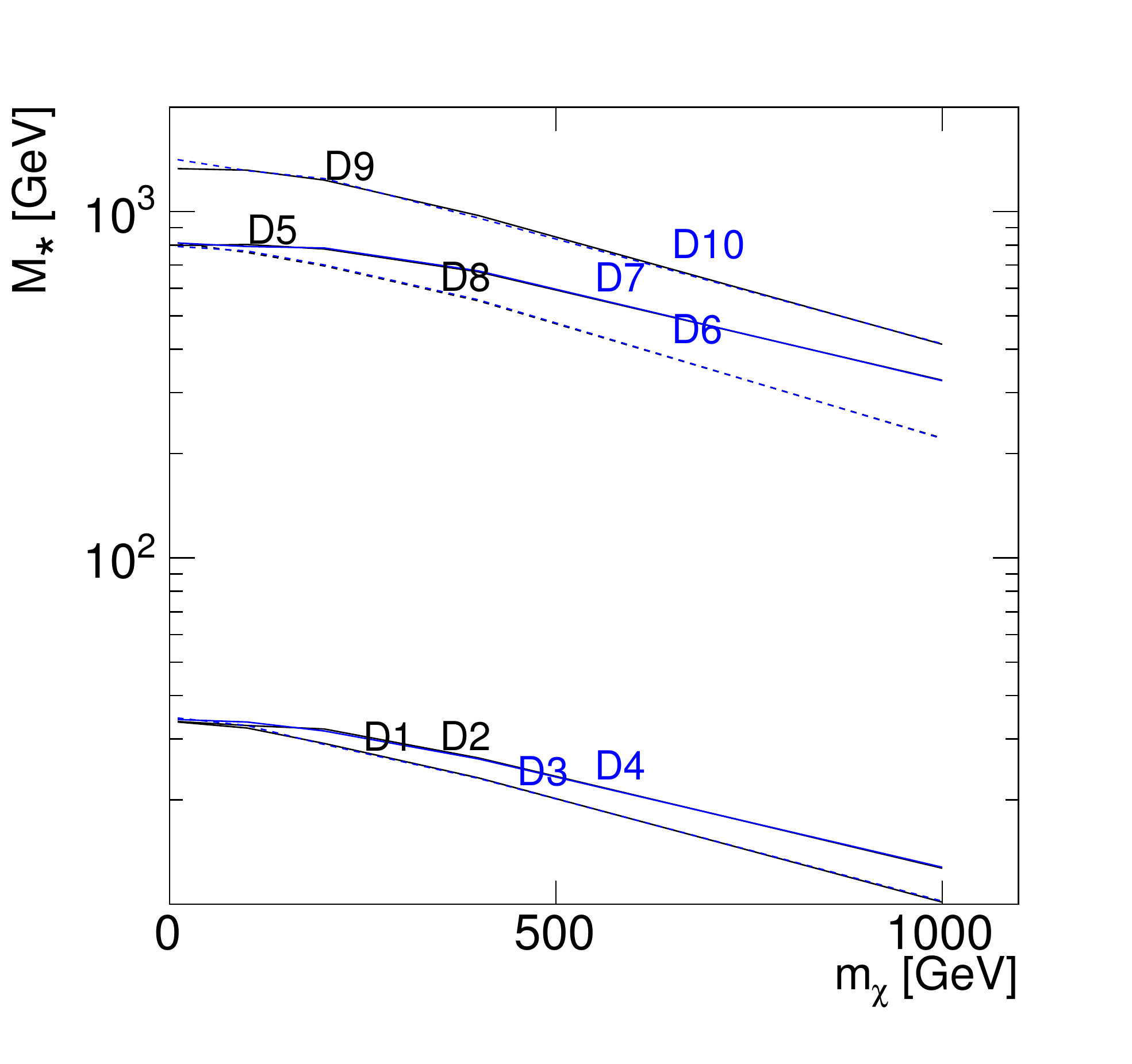}
\includegraphics[width=2.5in]{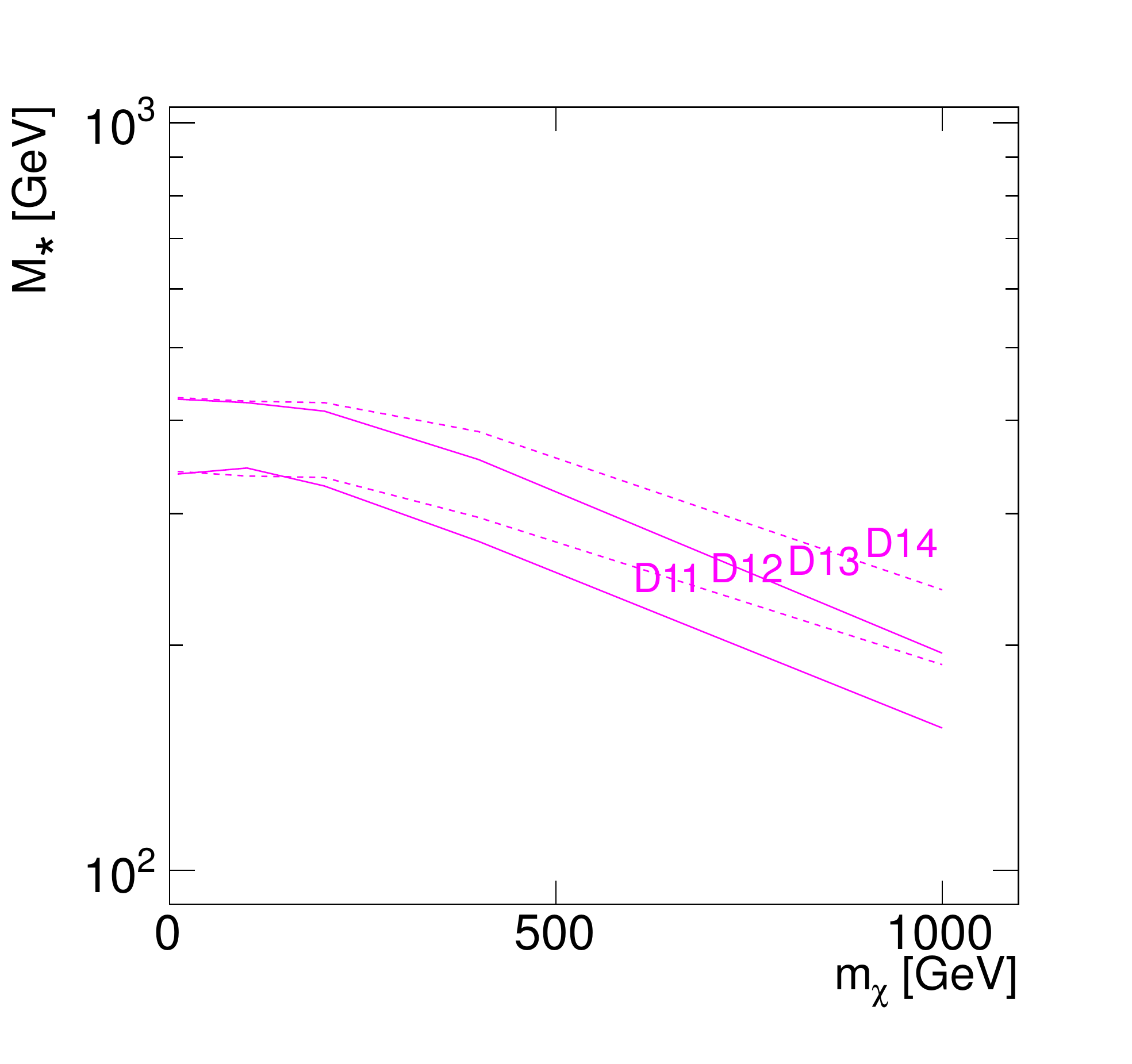}
\includegraphics[width=2.5in]{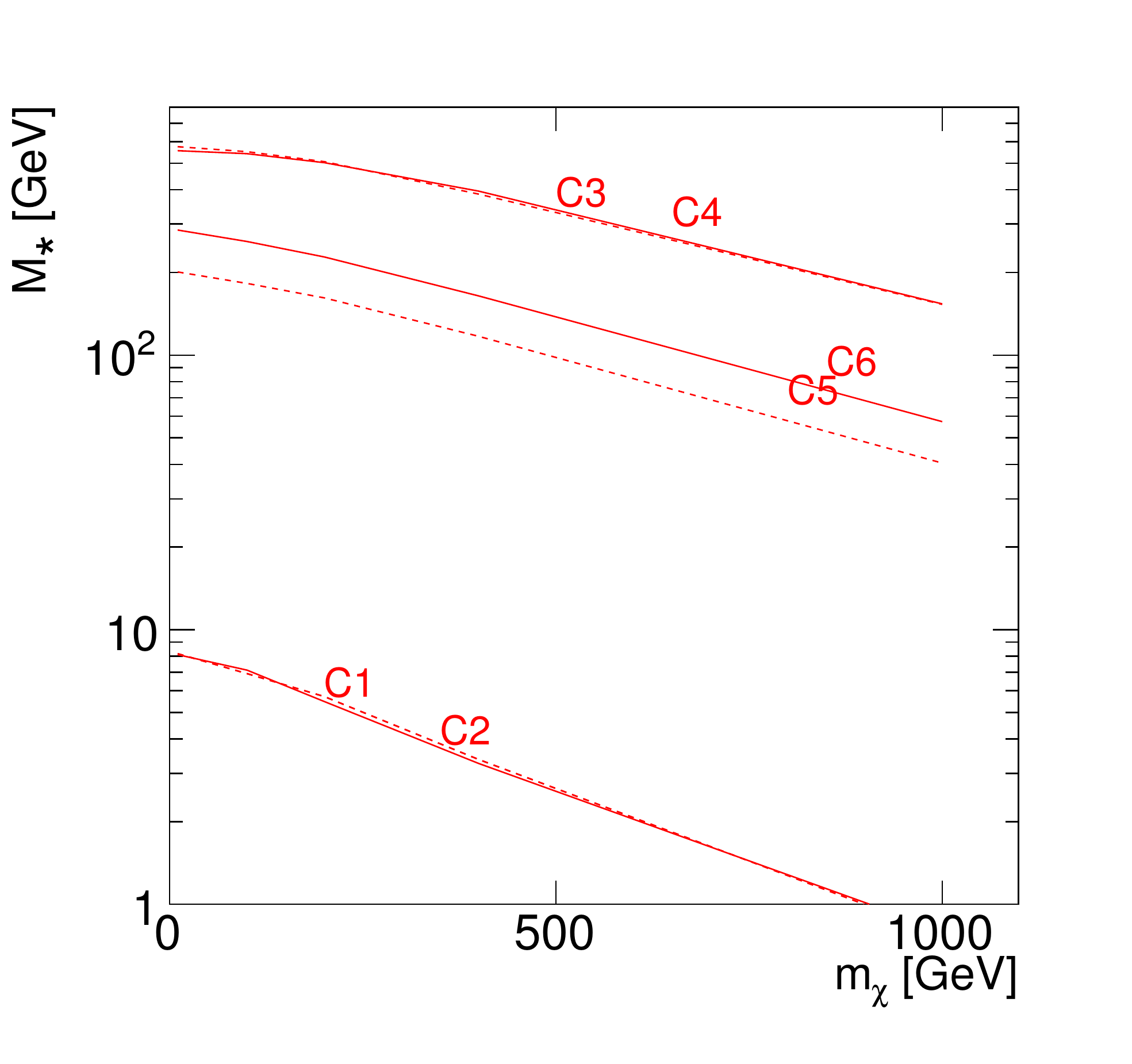}
\caption{ Combined limits on $M_\star$ at 90\% CL, using all available channels, for operators
  D1-14 and C1-C5 as a function of $m_\chi$.}
\label{fig:comb2}
\end{figure}

\begin{figure}
\includegraphics[width=3in]{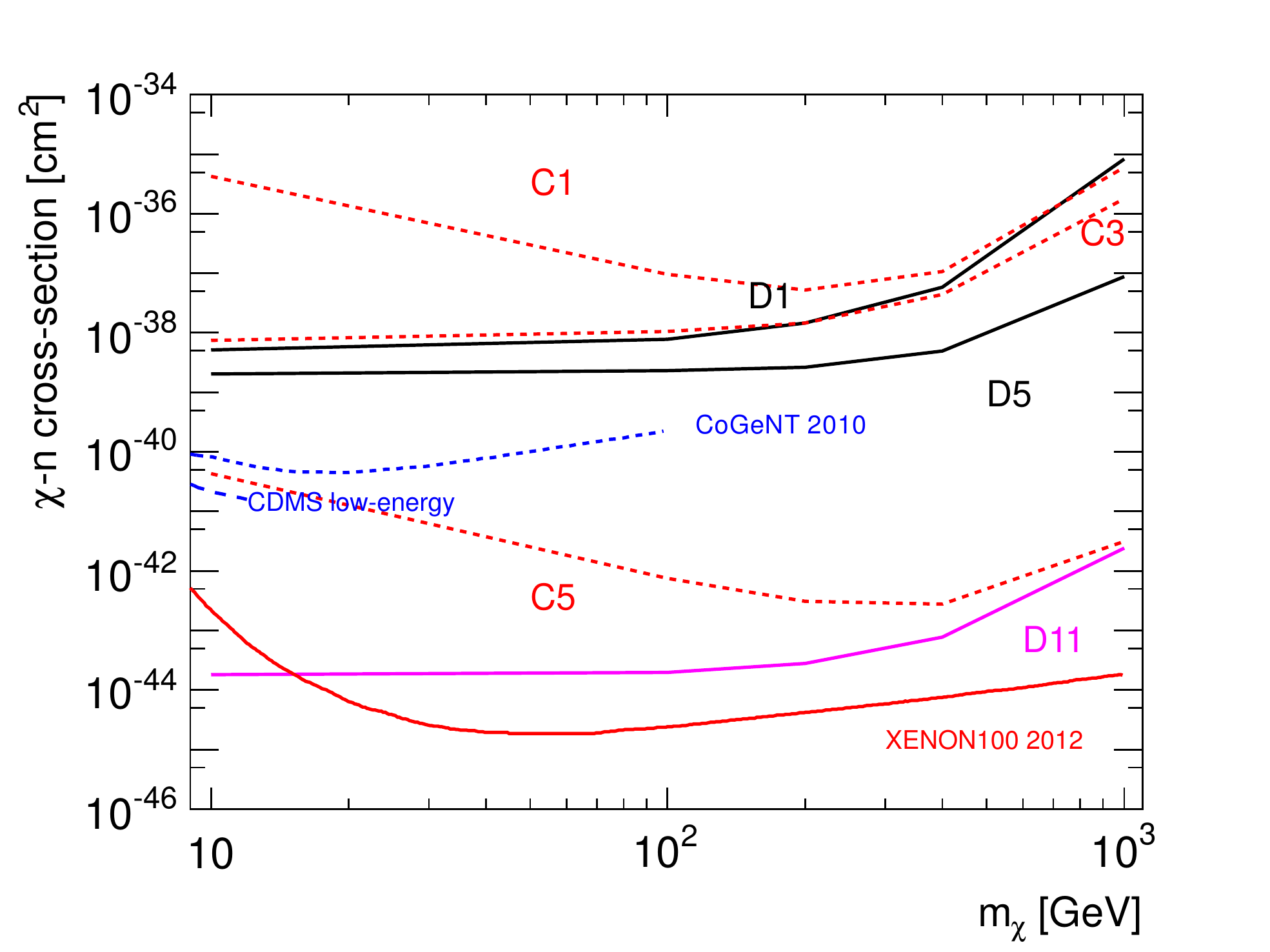}
\includegraphics[width=3in]{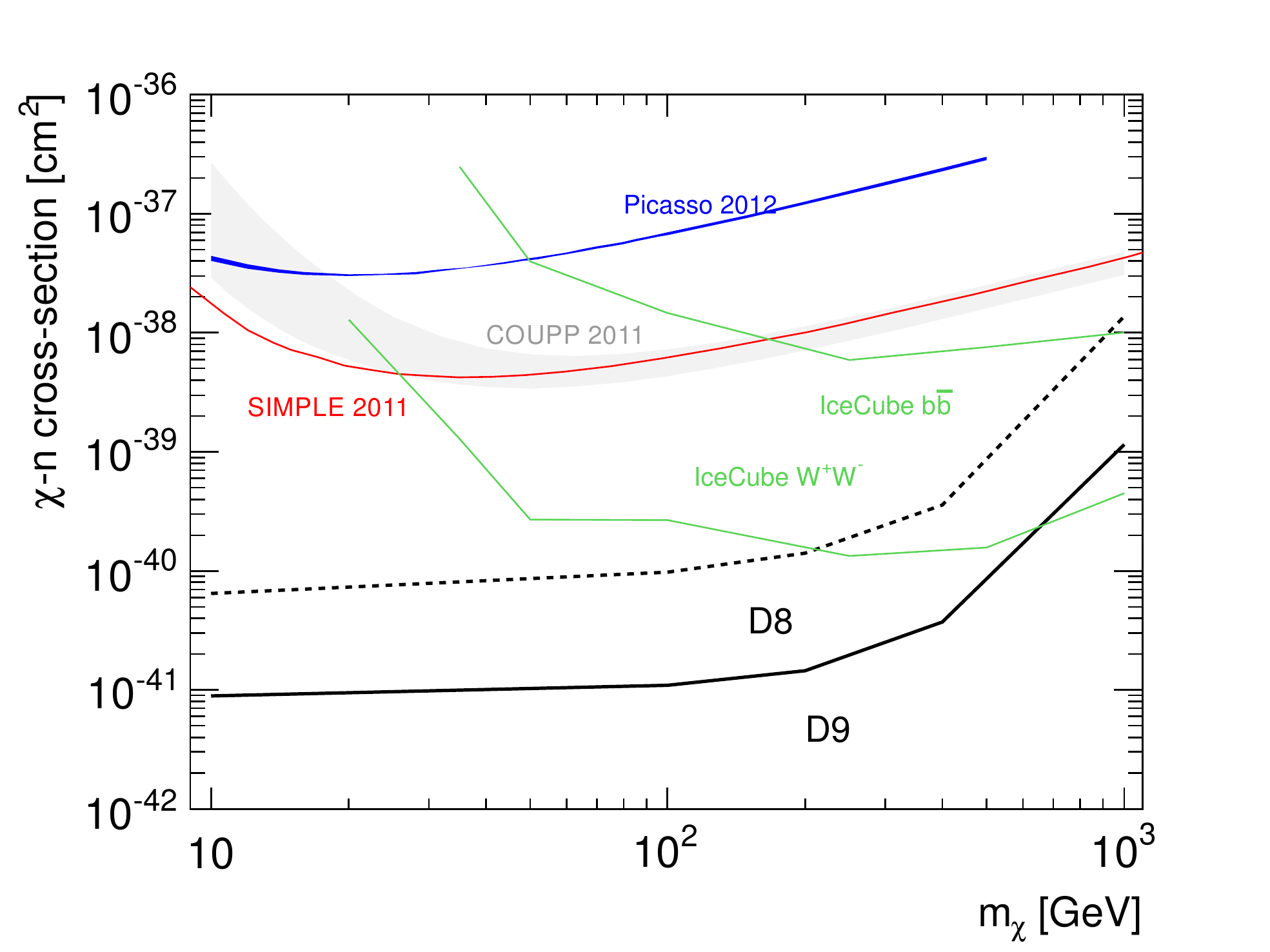}
\includegraphics[width=3in]{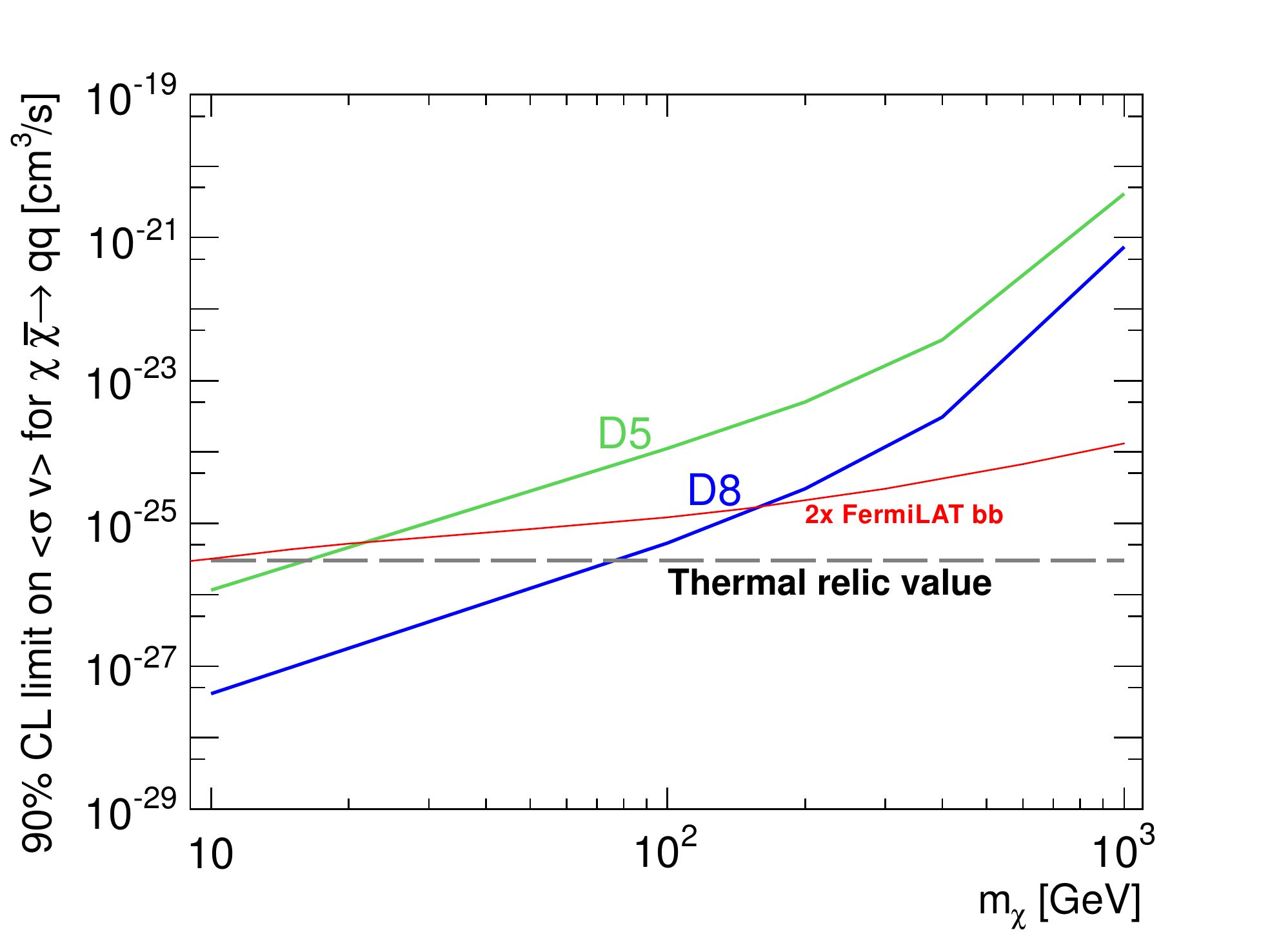}
\caption{ Top and center, limits at 90\% CL on the spin-independent
  and spin-dependent WIMP-nucleon cross section, $\sigma(\chi -n)$,
  for available operators. Bottom, interpretation of the limits on D5
  and D8 in terms of the velocity-averaged WIMP-annhiliation cross
  section, as defined in Ref~\cite{Fox:2011pm}.}
\label{fig:interp}
\end{figure}

We re-interpret the experimental analyses in the context of each
operator and perform the grand combination across all channels.
Figure~\ref{fig:comb2} and Table~\ref{tab:comb} show the limits on $M_\star$,
translated to the WIMP-nucleon cross section where possible.  In
addition, we translate the limits on D5 and D8 into limits on the WIMP
annihilation cross section, see Fig.~\ref{fig:interp}.

\begin{table}
\caption{Combined limits on $M_\star$ at 90\% CL, using all available channels, for operators
  D1-11 and C1-C5 for low and high values of the WIMP mass
  $m_\chi$. Where possible, limits are shown on the WIMP-nucleon cross
  section, $\sigma(\chi -n)$.}
\label{tab:comb}
\begin{tabular}{lrrr}
\hline
\hline
Operator & $m_{\chi}$ & $M_\star$ & $\sigma(\chi-n)$ \\
 & (GeV) & (GeV) & [cm$^{2}$] \\
\hline 
D1        & 10    &  34   & 5.2$\times 10^{-39}$\\ 
         & 1000  &  10   & 8.3$\times 10^{-36}$\\ 
\hline 
D2        & 10    &  34   & \\ 
         & 1000  &  13   & \\ 
\hline 
D3        & 10    &  34   & \\ 
         & 1000  &  10   & \\ 
\hline 
D4        & 10    &  34   & \\ 
         & 1000  &  13   & \\ 
\hline 
D5        & 10    & 795   & 2.0$\times 10^{-39}$\\ 
         & 1000  & 325   & 8.8$\times 10^{-38}$\\ 
\hline 
D6        & 10    & 791   & \\ 
         & 1000  & 221   & \\ 
\hline 
D7        & 10    & 812   & \\ 
         & 1000  & 324   & \\ 
\hline 
D8        & 10    & 811   & 6.5$\times 10^{-41}$\\ 
         & 1000  & 222   & 1.4$\times 10^{-38}$\\ 
\hline 
D9        & 10    & 1331  & 8.9$\times 10^{-42}$\\ 
         & 1000  & 413   & 1.1$\times 10^{-39}$\\ 
\hline 
D10       & 10    & 1410  & \\ 
         & 1000  & 415   & \\ 
\hline 
D11       & 10    & 339   & 1.8$\times 10^{-44}$\\ 
         & 1000  & 155   & 2.4$\times 10^{-42}$\\ 
\hline 
D12       & 10    & 342   & \\ 
         & 1000  & 188   & \\ 
\hline 
D13       & 10    & 427   & \\ 
         & 1000  & 195   & \\ 
\hline 
D14       & 10    & 429   & \\ 
         & 1000  & 237  & \\ 
\hline 
C1        & 10    &   8   & 4.2$\times 10^{-36}$\\ 
         & 1000  &   1   & 6.1$\times 10^{-36}$\\ 
\hline 
C2        & 10    &   8   & \\ 
         & 1000  &   1   & \\ 
\hline 
C3        & 10    & 575   & 7.5$\times 10^{-39}$\\ 
         & 1000  & 153   & 1.8$\times 10^{-36}$\\ 
\hline 
C4        & 10    & 556   & \\ 
         & 1000  & 154   & \\ 
\hline 
C5        & 10    & 201   & 4.4$\times 10^{-41}$\\ 
         & 1000  &  41   & 3.1$\times 10^{-42}$\\ 
\hline 
C6        & 10    & 286   &\\ 
         & 1000  &  57   &\\ 
\hline
\hline
\end{tabular}

\end{table}

\subsection{Conclusions}

We have presented the first combination of collider-based searches for
dark matter pair production, using final states involving jets,
photons and leptonically-decaying $Z$ bosons in the context of
effective field theories.  The most powerful results are from the mono-jet
analyses, and the greatest gains come from the combination of the
independent analyses from ATLAS and CMS, though the other final states
make a non-negligble improvement. The results are the strongest limits
to date from collider searches in the effective field theory context.

In addition, we have reinterpreted the experimental results, quoted by
ATLAS and CMS only for a few effective operators, across a broad range
of operators, providing a comprehensive view of the power of these
searches to constrain the weak-level or weaker interactions between
dark matter and standard model particles.

We have made use of the effective field theory framework to convert
the ATLAS and CMS results to quantities relevant for direct-detection
and indirect-detection dark matter searches. Under the assumptions
made for the effective operators, LHC limits can be very competitive,
in particular for low-mass dark matter particles $m_\chi \le 10$~GeV.

\subsection{Acknowledgements}

We acknowledge useful conversations with Tim Tait, Roni Harnik and Patrick Fox.
DW and NZ are supported by grants from the Department of Energy
Office of Science and by the Alfred P. Sloan Foundation.

\clearpage
\appendix

\section{Appendix: Individual Operators}

\begin{figure}[htb!]
\includegraphics[width=2in]{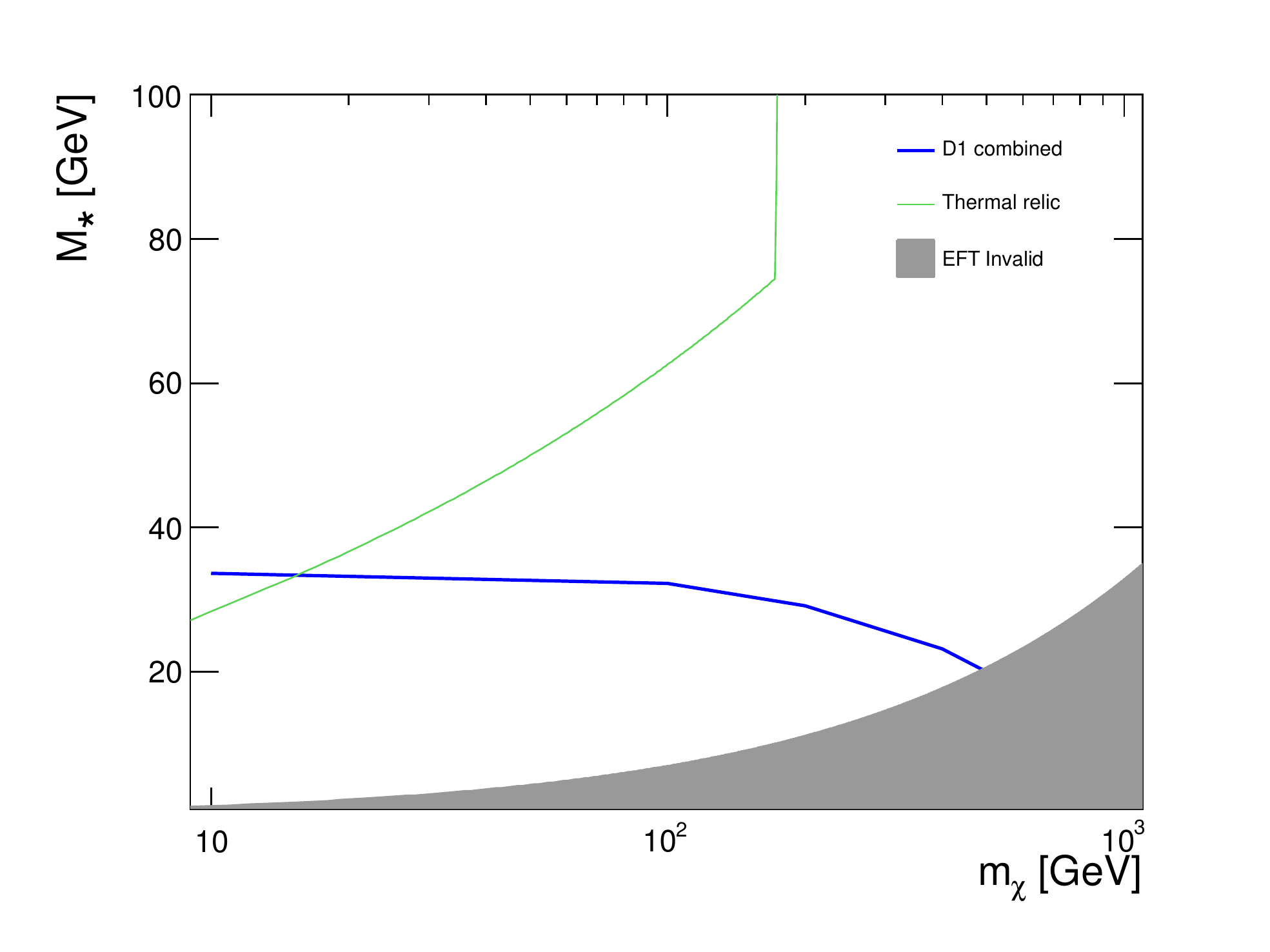}
\includegraphics[width=2in]{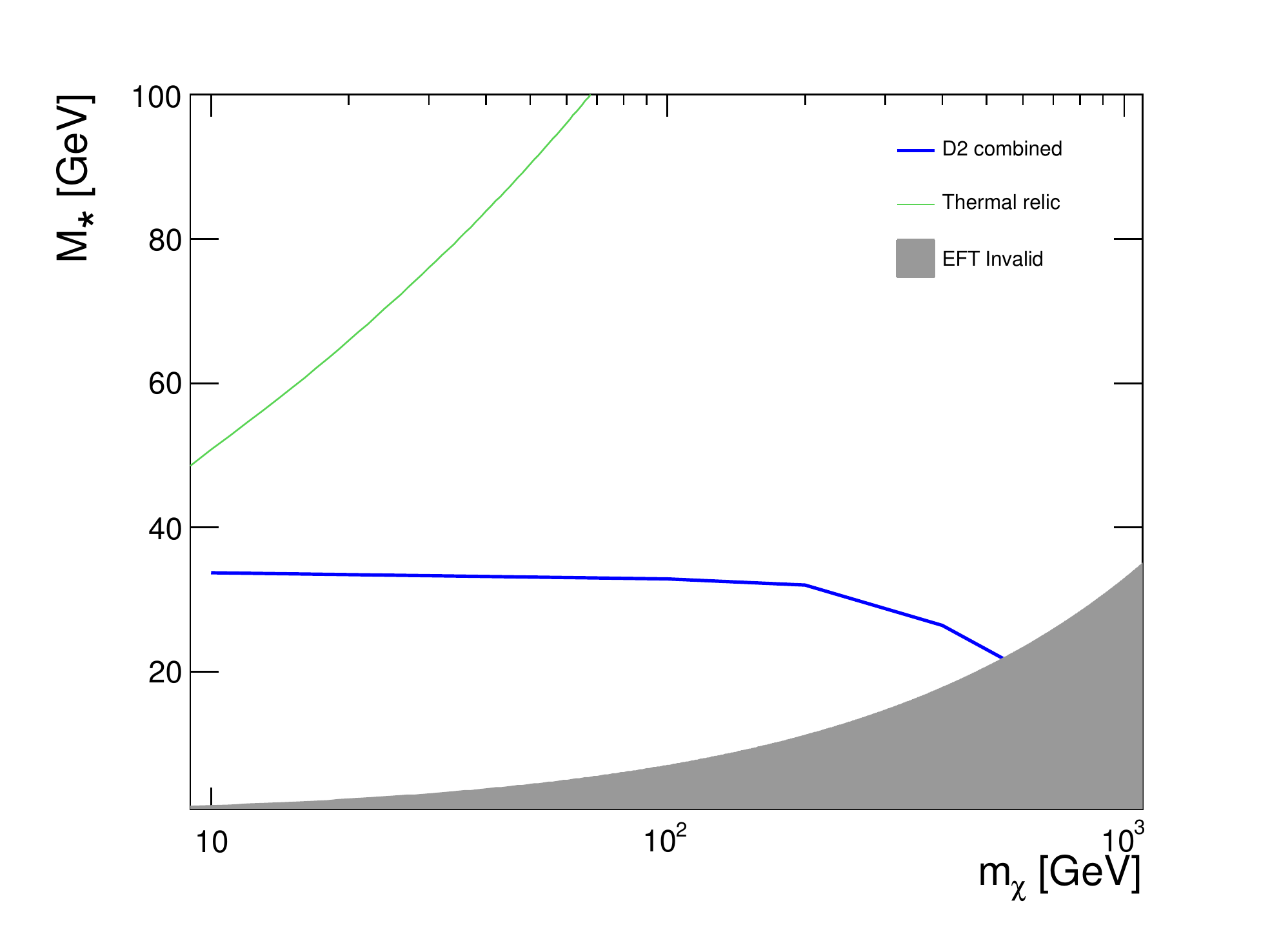}
\includegraphics[width=2in]{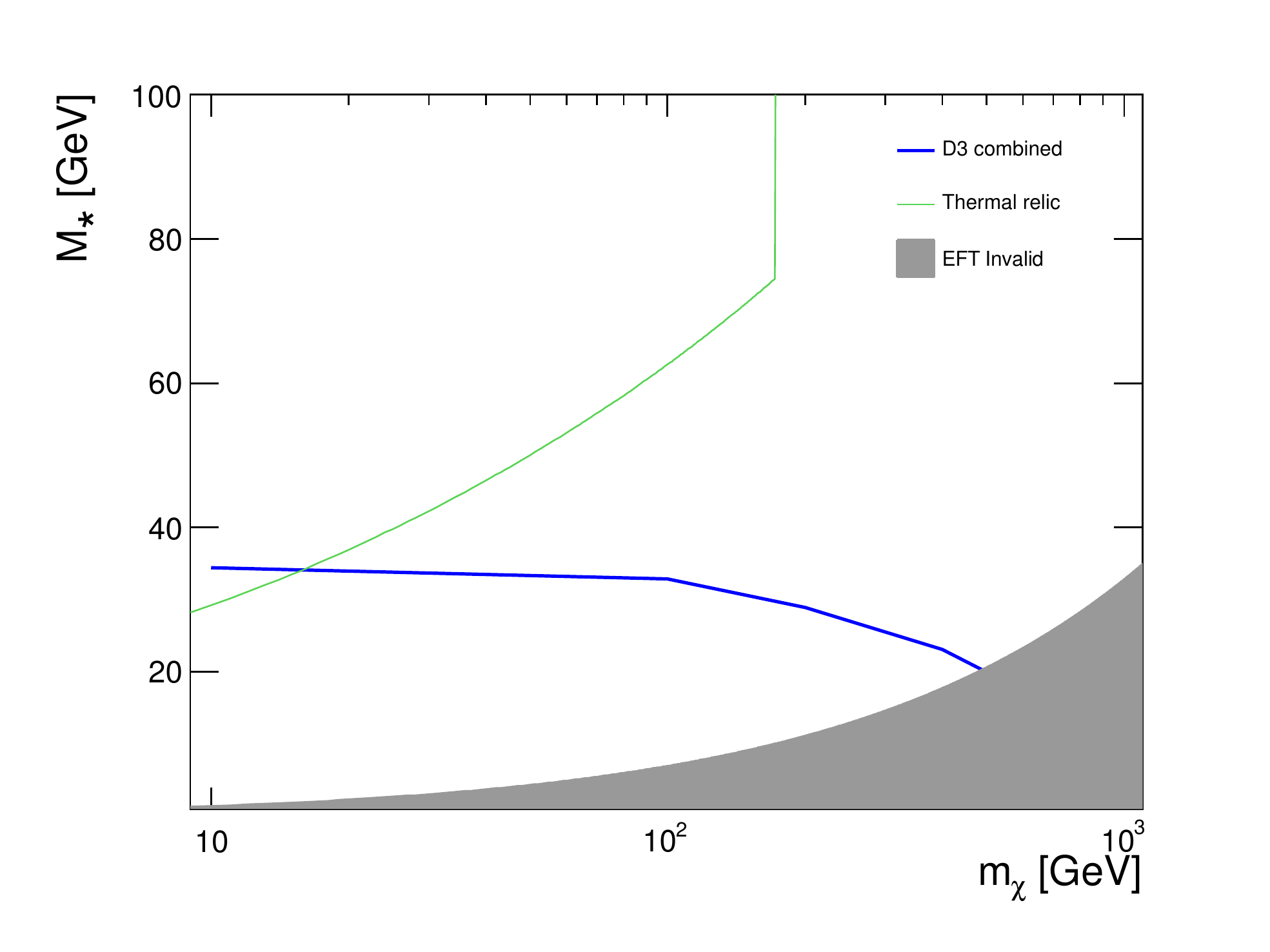}
\includegraphics[width=2in]{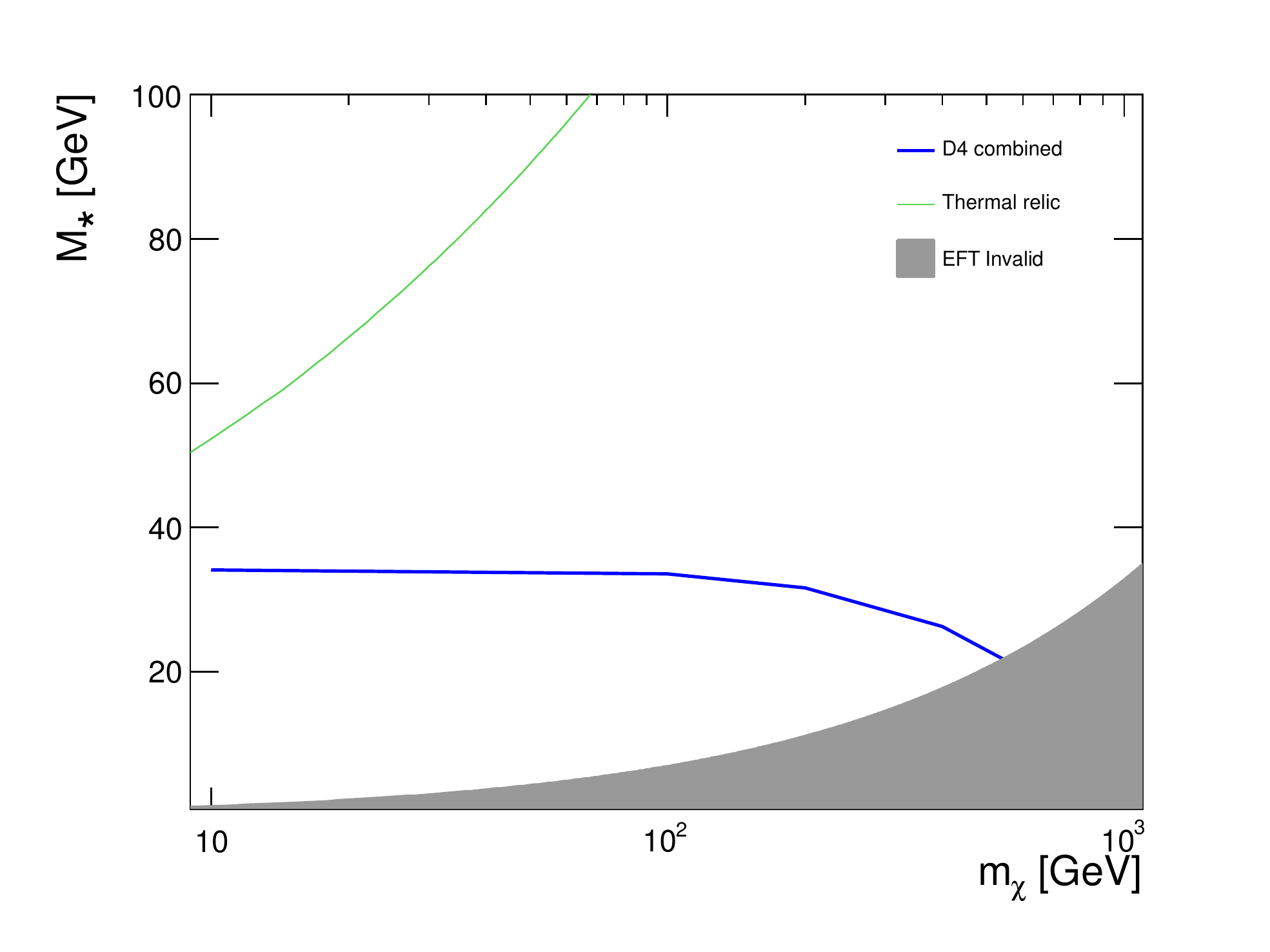}
\caption{ Combined limits on $M_\star$ versus dark matter mass
  $m_\chi$ for operators D1, D2, D3 and D4. The $M_\star$ values at which dark matter particles of a given
  mass would result in the required relic abundance are shown as
  green lines~\cite{Goodman:2010ku}, assuming annihilation in the early universe
  proceeded exclusively via the given operator. }
\label{fig:simple1}
\end{figure}

\begin{figure}[htb!]
\includegraphics[width=2in]{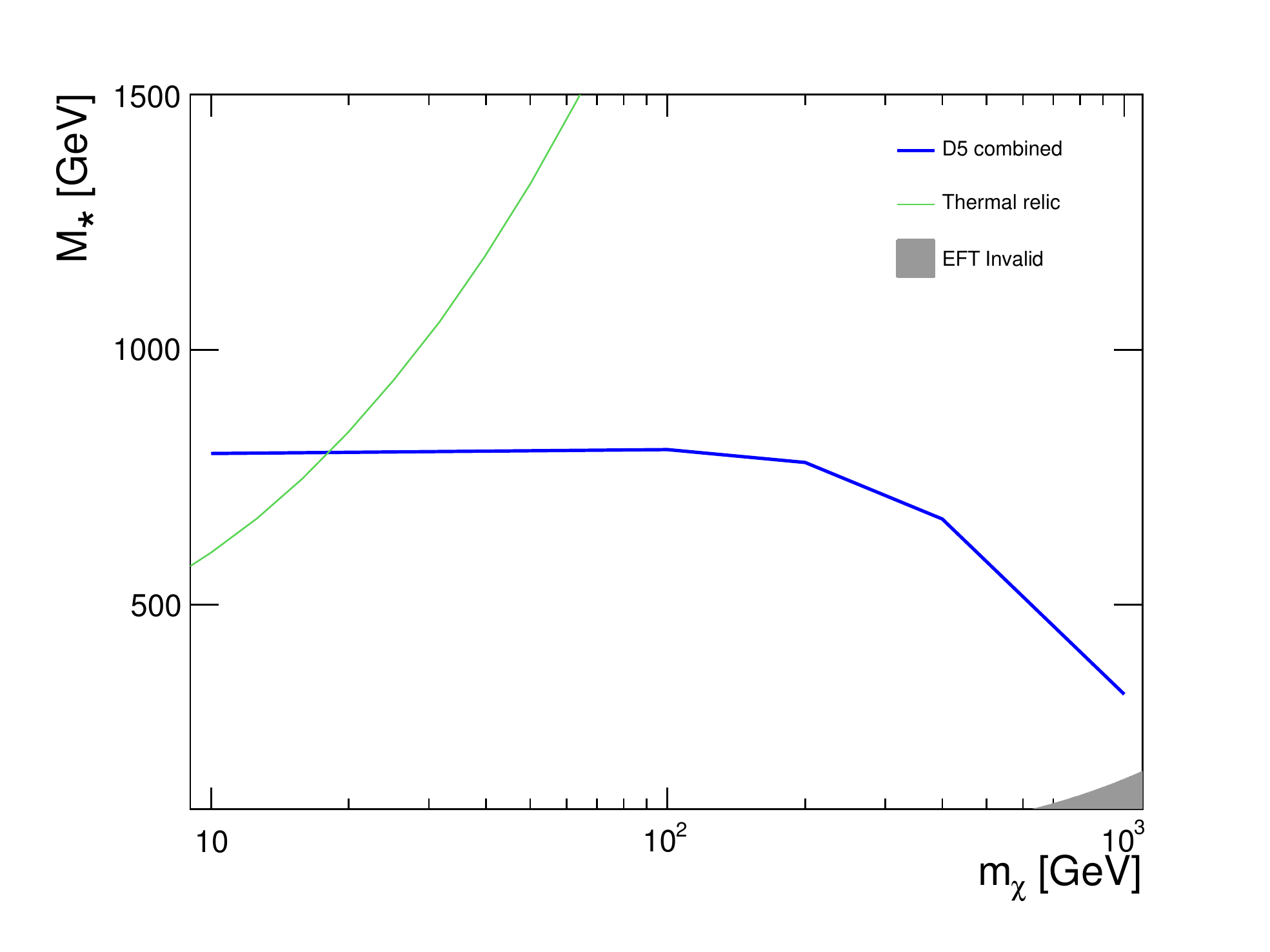}
\includegraphics[width=2in]{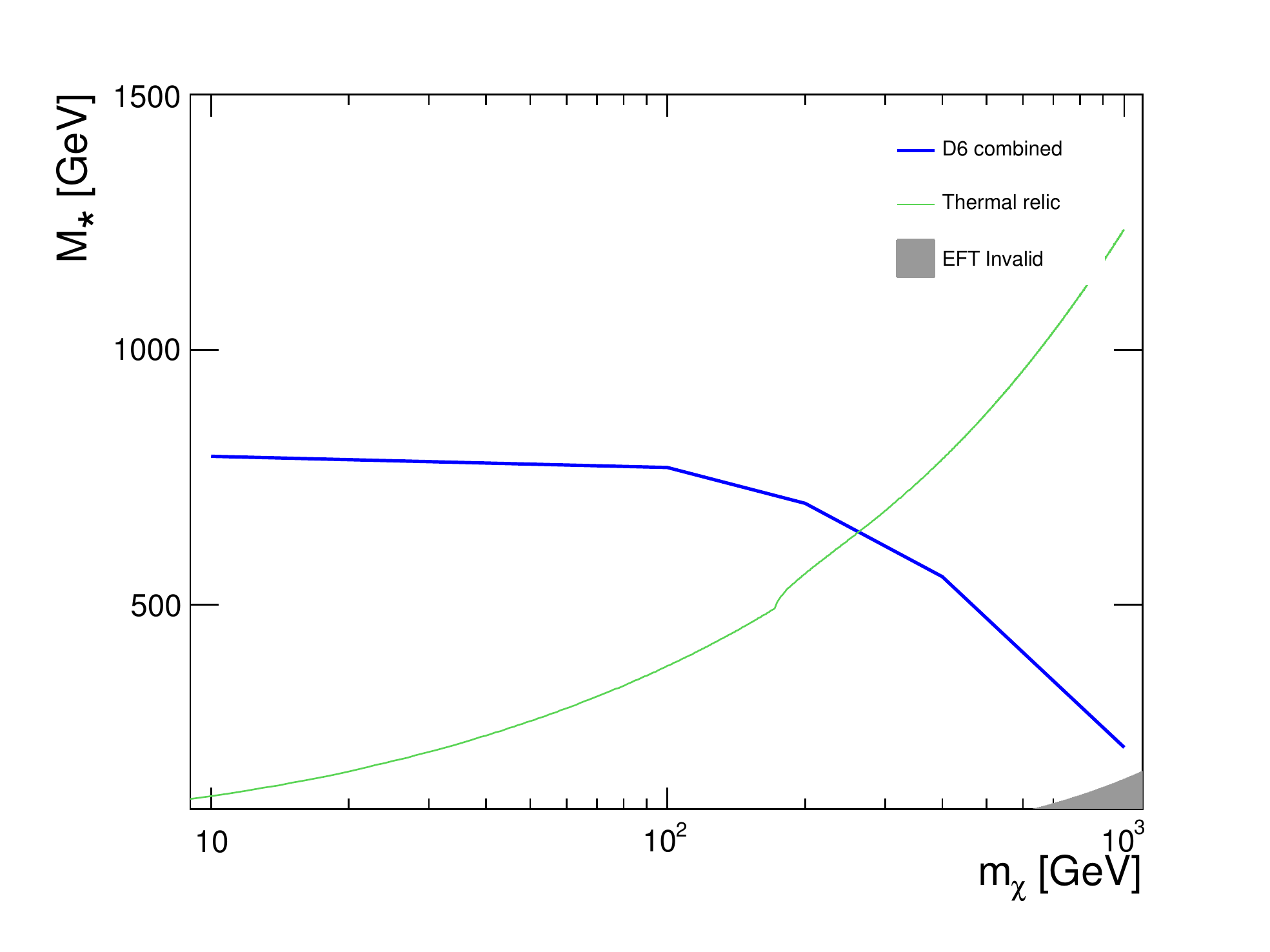}
\includegraphics[width=2in]{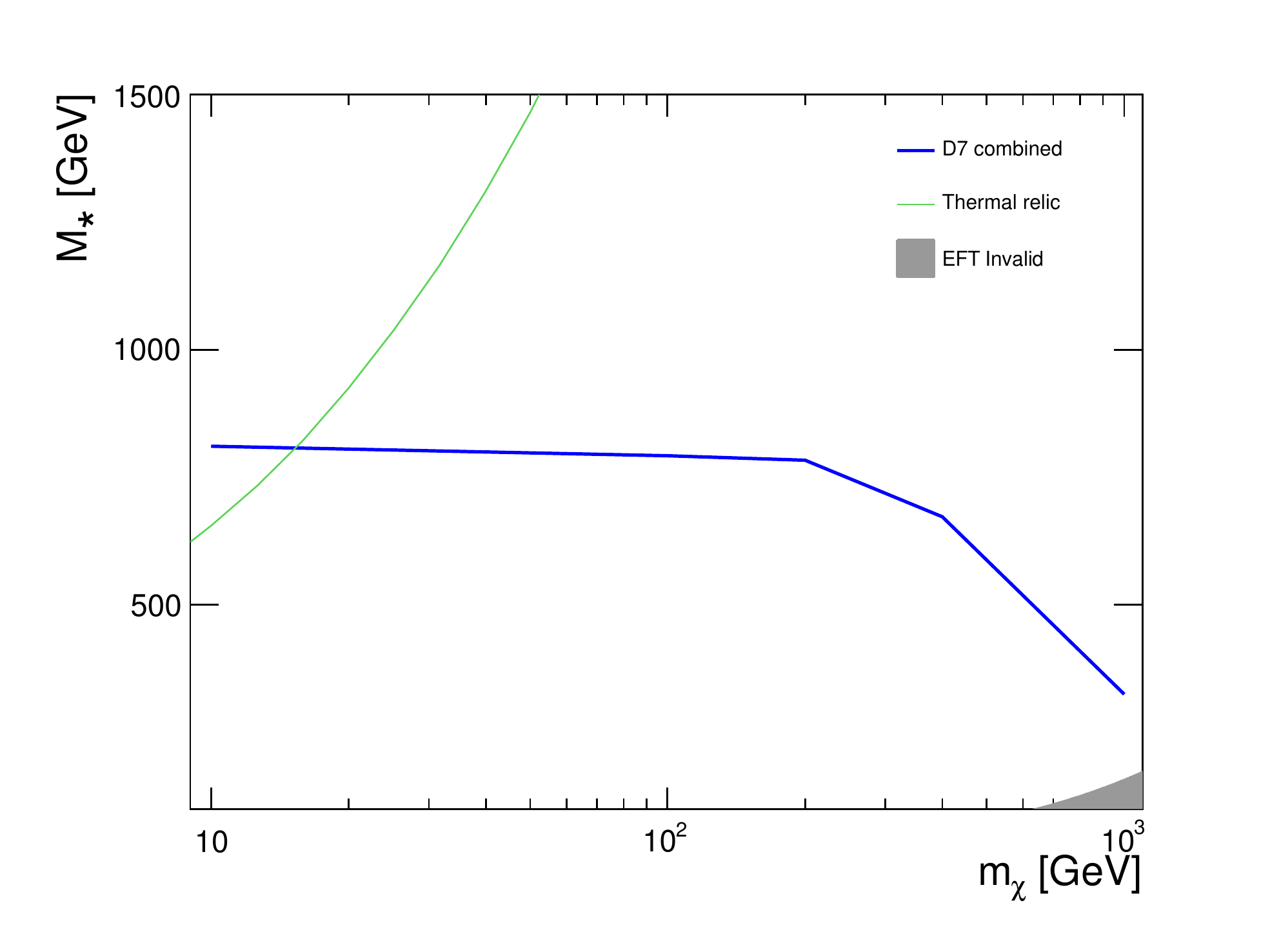}
\caption{ Combined limits on $M_\star$ versus dark matter mass
  $m_\chi$ for operators D5, D6, and D7. The $M_\star$ values at which dark matter particles of a given
  mass would result in the required relic abundance are shown as
  green lines~\cite{Goodman:2010ku}, assuming annihilation in the early universe
  proceeded exclusively via the given operator.}
\label{fig:simple2a}
\end{figure}

\begin{figure}[htb!]
\includegraphics[width=2in]{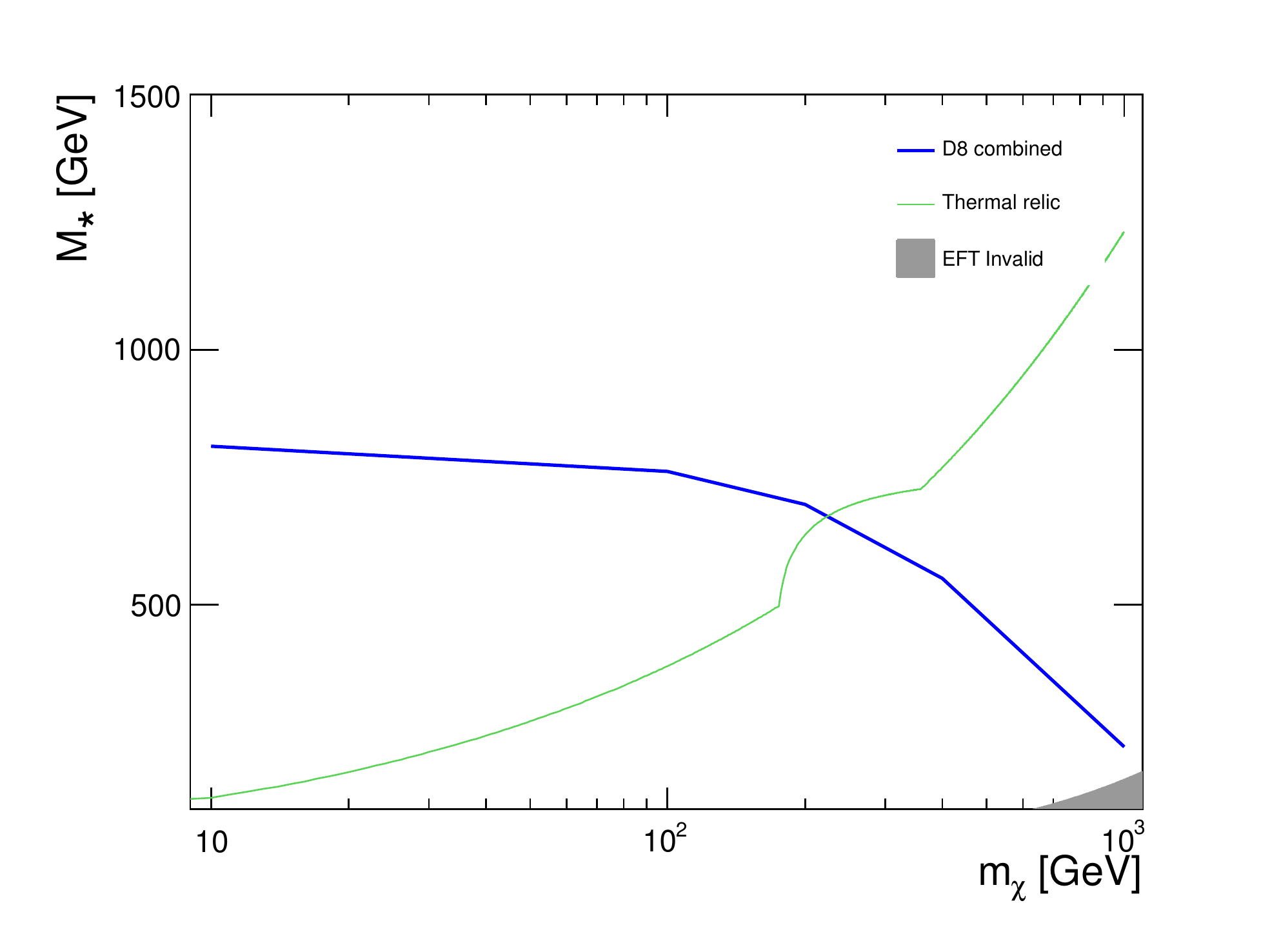}
\includegraphics[width=2in]{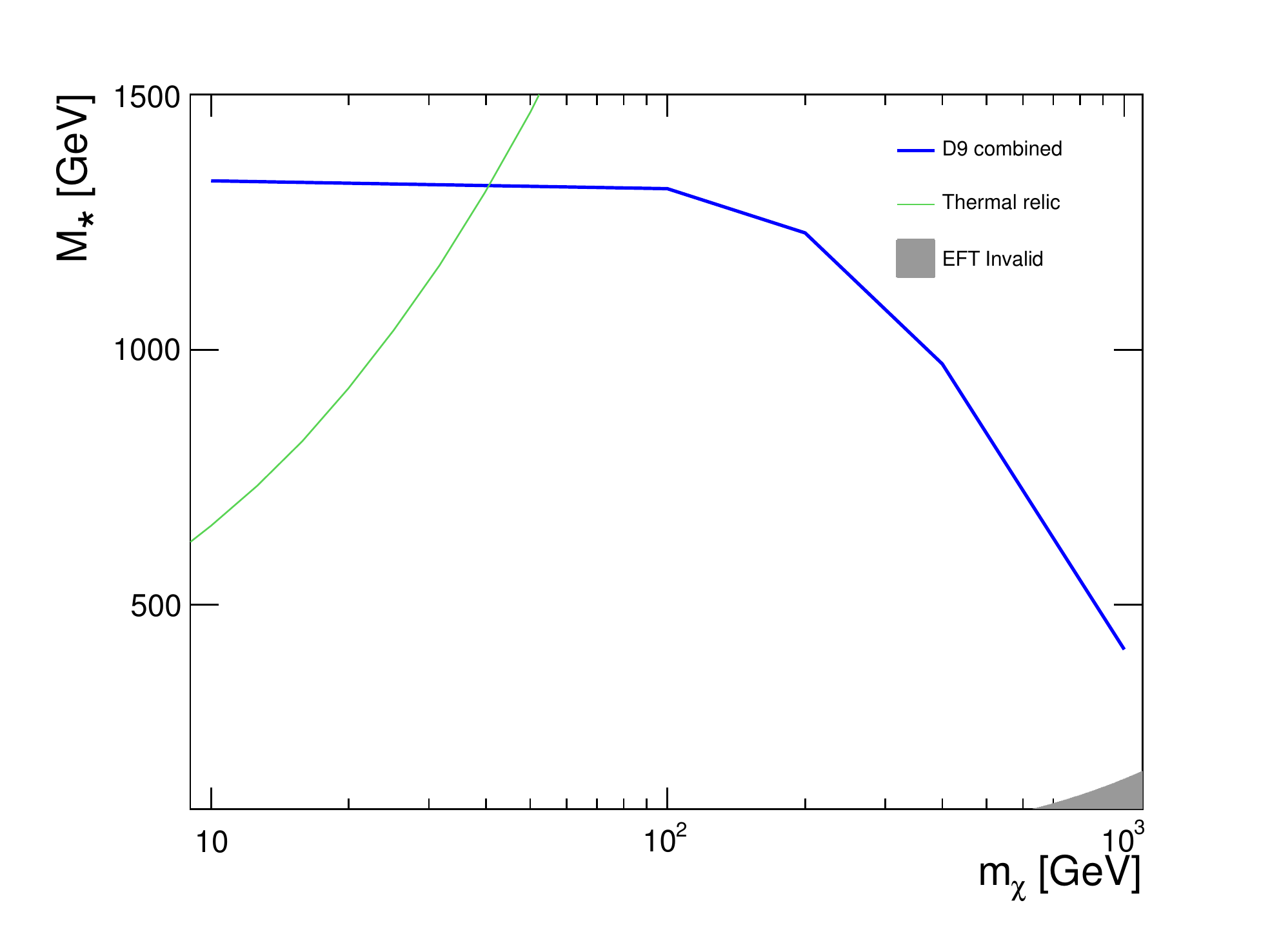}
\includegraphics[width=2in]{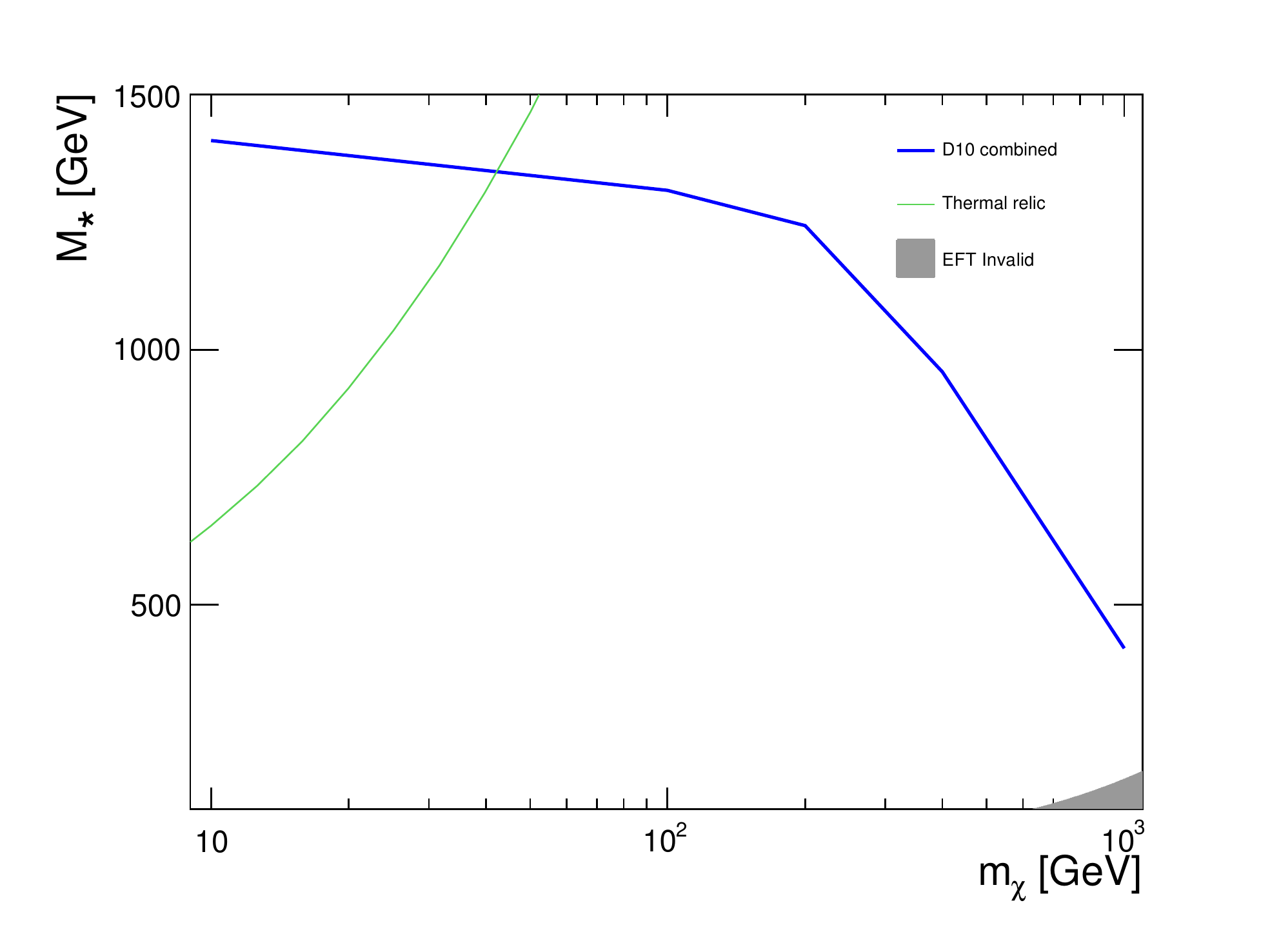}
\caption{ Combined limits on $M_\star$ versus dark matter mass
  $m_\chi$ for operators D8,D9 and D10. The $M_\star$ values at which dark matter particles of a given
  mass would result in the required relic abundance are shown as
  green lines~\cite{Goodman:2010ku}, assuming annihilation in the early universe
  proceeded exclusively via the given operator.}
\label{fig:simple2b}
\end{figure}

\begin{figure}[htb!]
\includegraphics[width=2in]{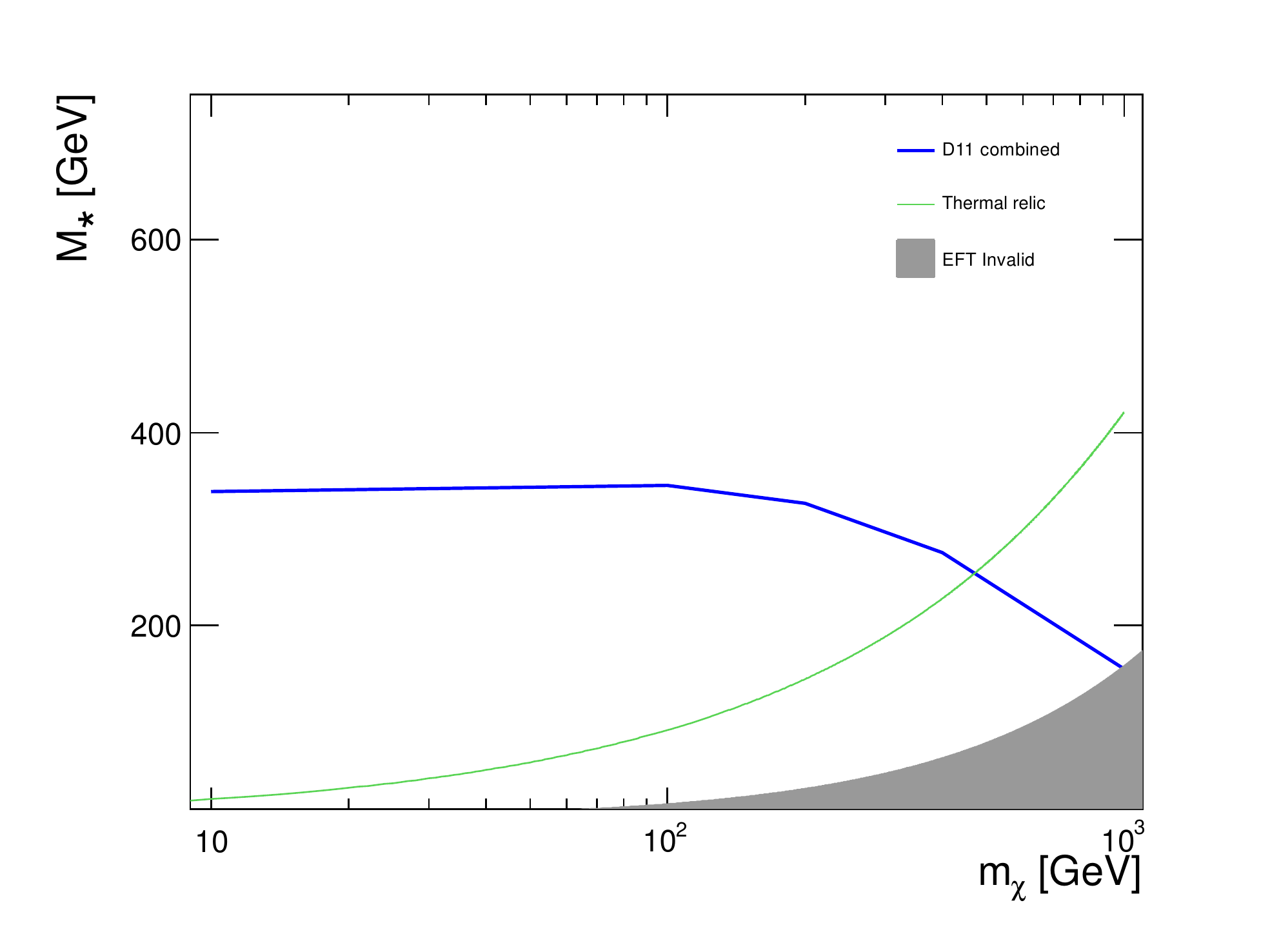}
\includegraphics[width=2in]{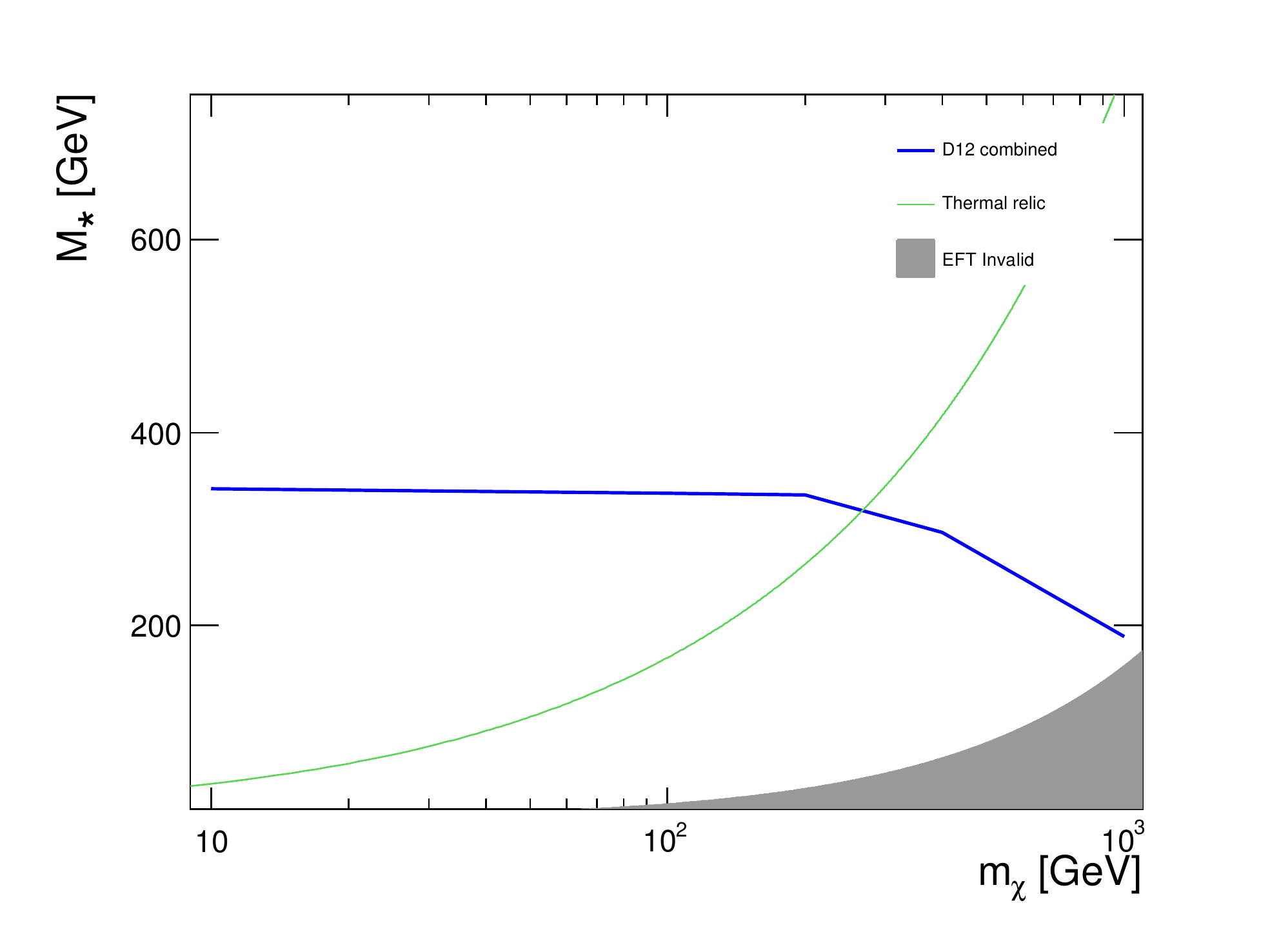}
\includegraphics[width=2in]{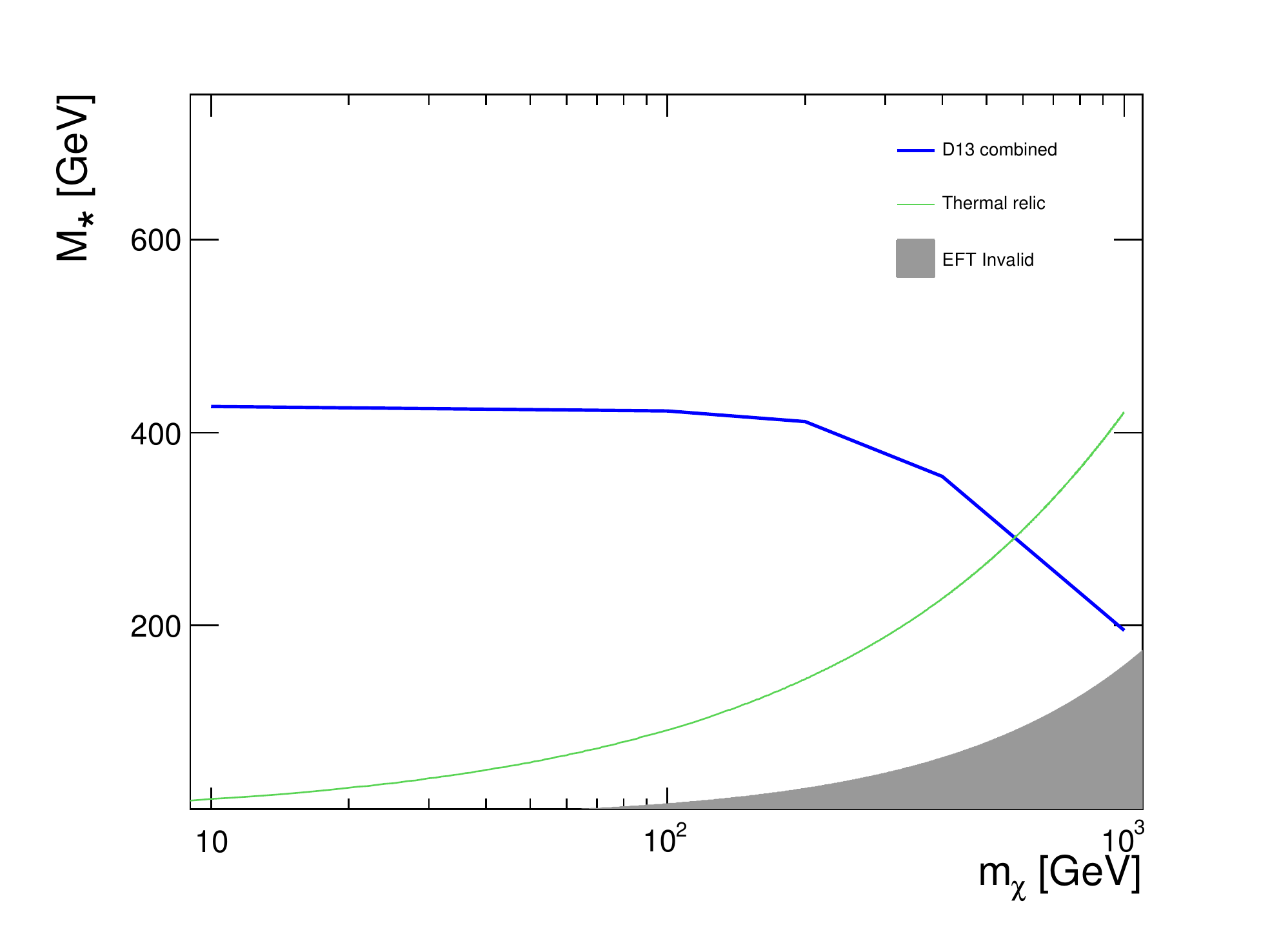}
\includegraphics[width=2in]{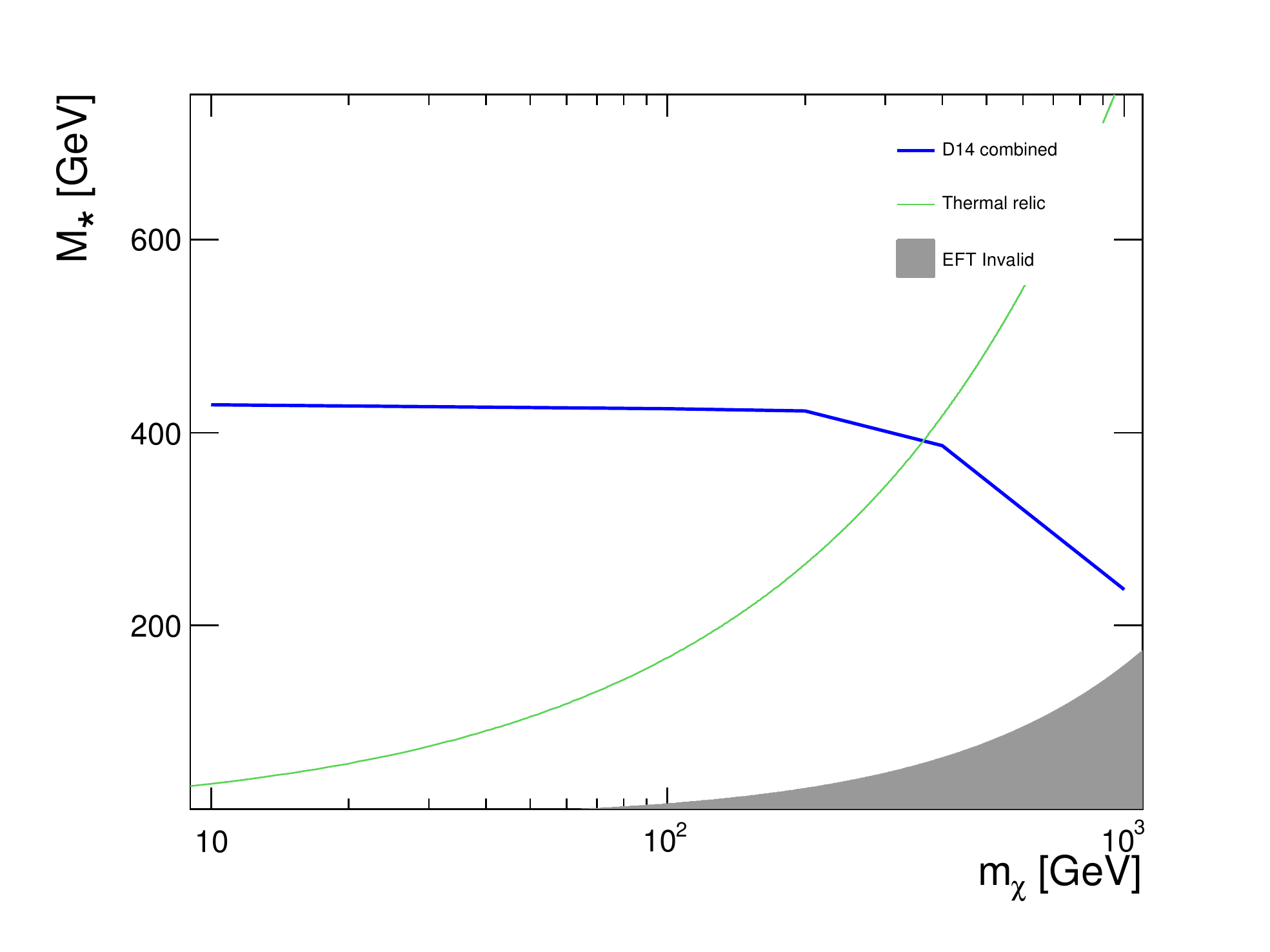}
\caption{ Combined limits on $M_\star$ versus dark matter mass
  $m_\chi$ for operators D11, D12, D13 and D14. The $M_\star$ values at which dark matter particles of a given
  mass would result in the required relic abundance are shown as
  green lines~\cite{Goodman:2010ku}, assuming annihilation in the early universe
  proceeded exclusively via the given operator.}
\label{fig:simple3}
\end{figure}

\begin{figure}[htb!]
\includegraphics[width=2in]{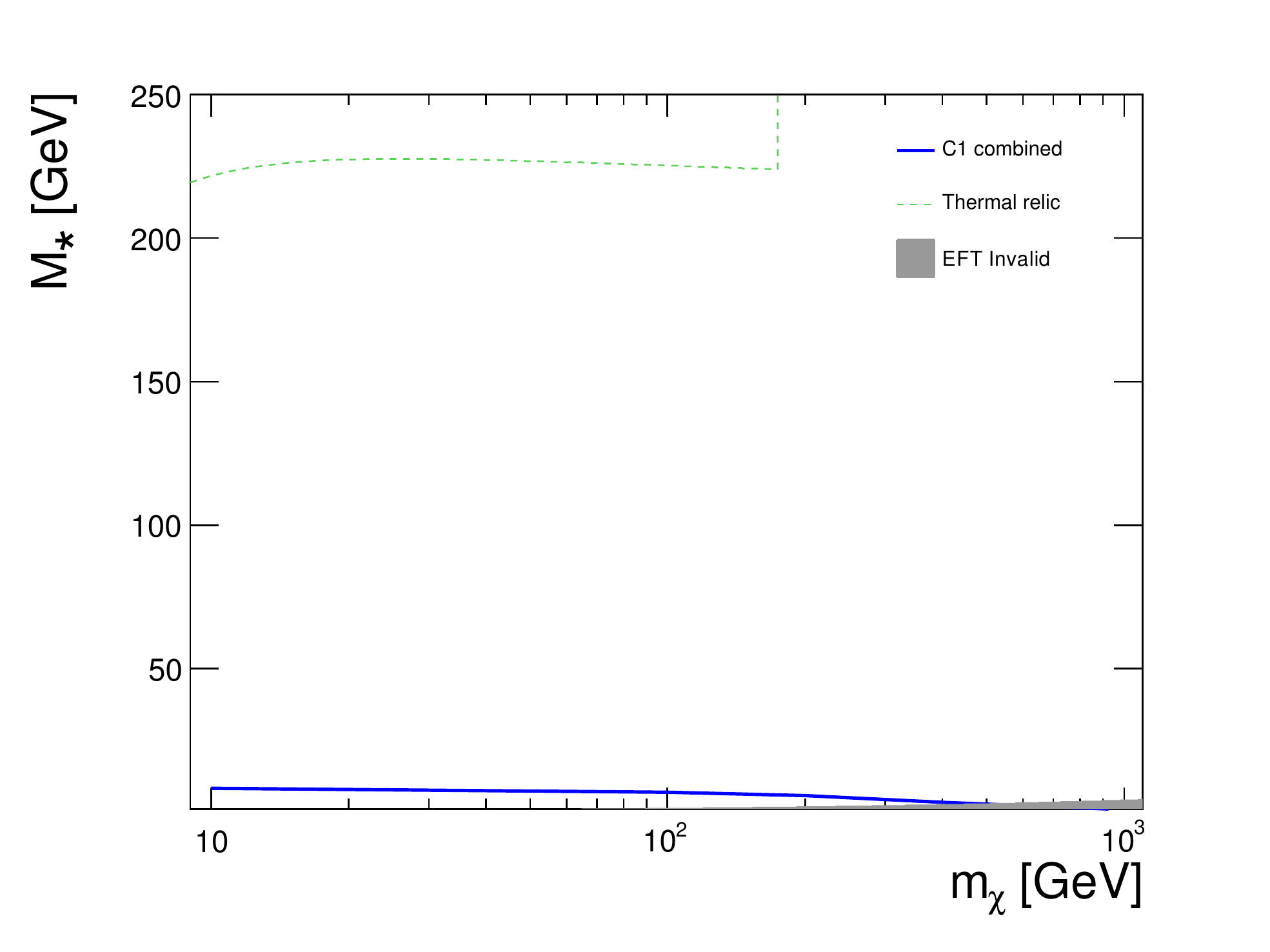}
\includegraphics[width=2in]{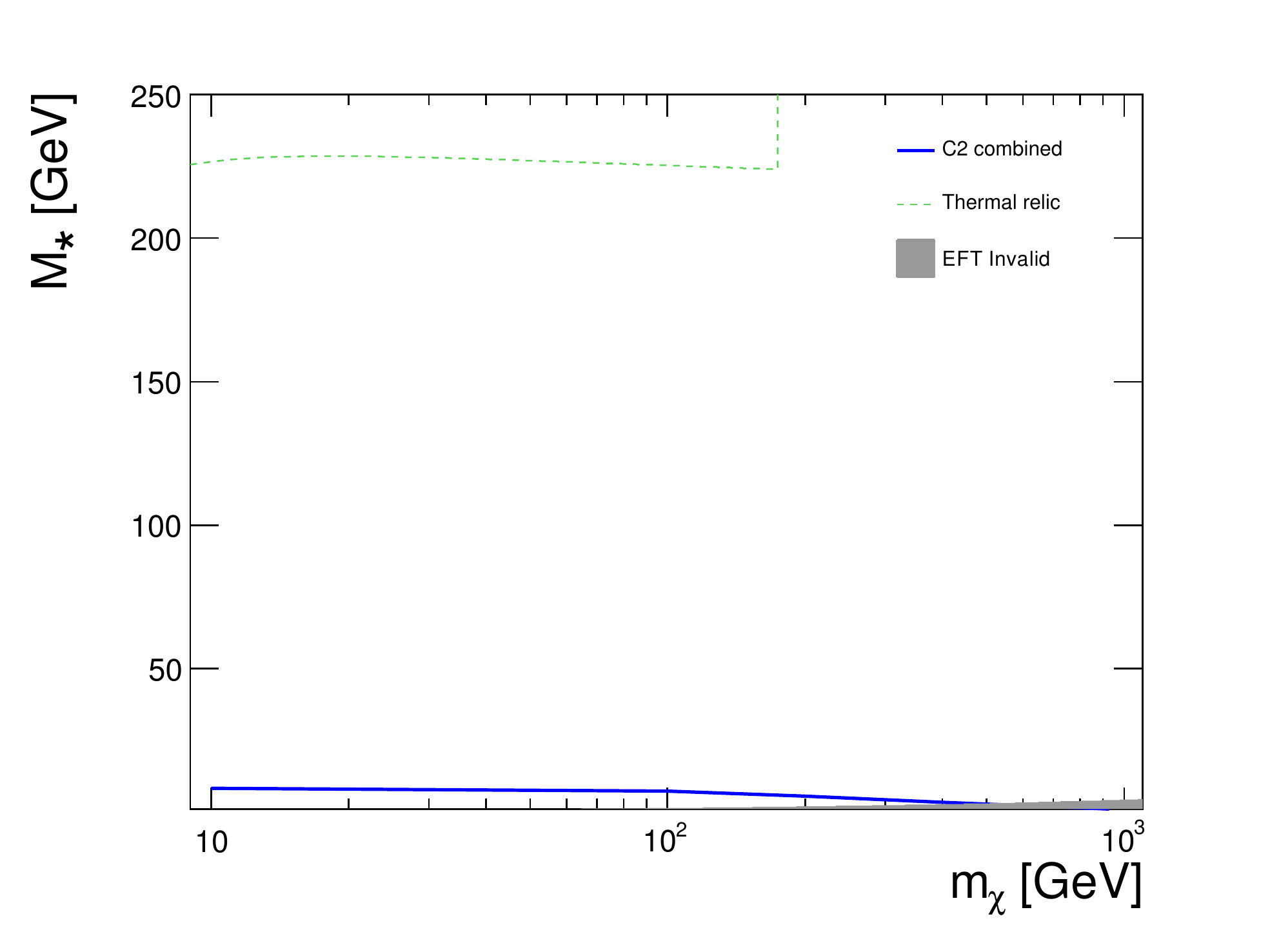}
\includegraphics[width=2in]{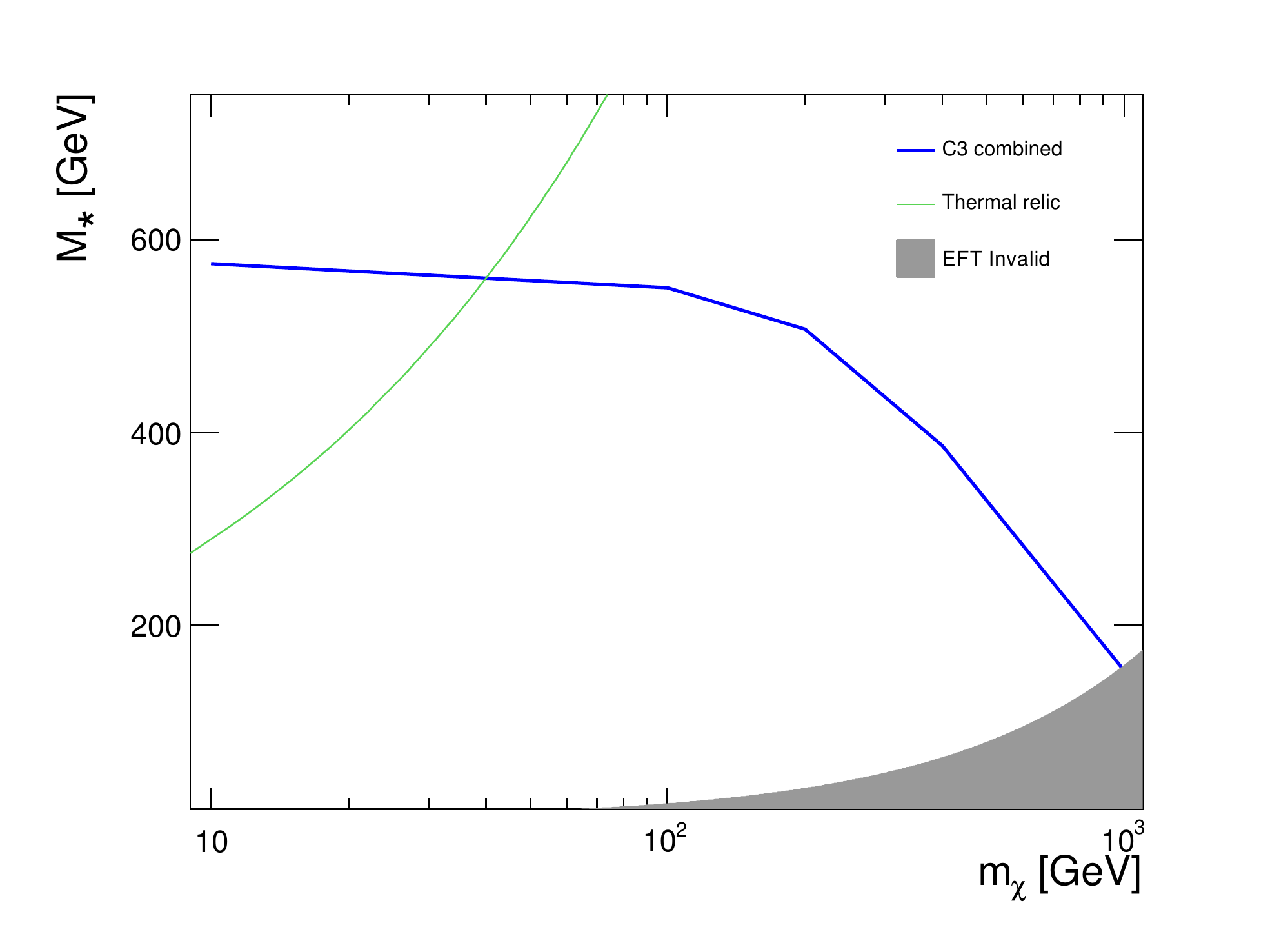}
\caption{ Combined limits on $M_\star$ versus dark matter mass
  $m_\chi$ for operators C1, C2, and C3. The $M_\star$ values at which dark matter particles of a given
  mass would result in the required relic abundance are shown as
  green lines~\cite{Goodman:2010ku}, assuming annihilation in the early universe
  proceeded exclusively via the given operator.}
\label{fig:simple4a}
\end{figure}

\begin{figure}[htb!]
\includegraphics[width=2in]{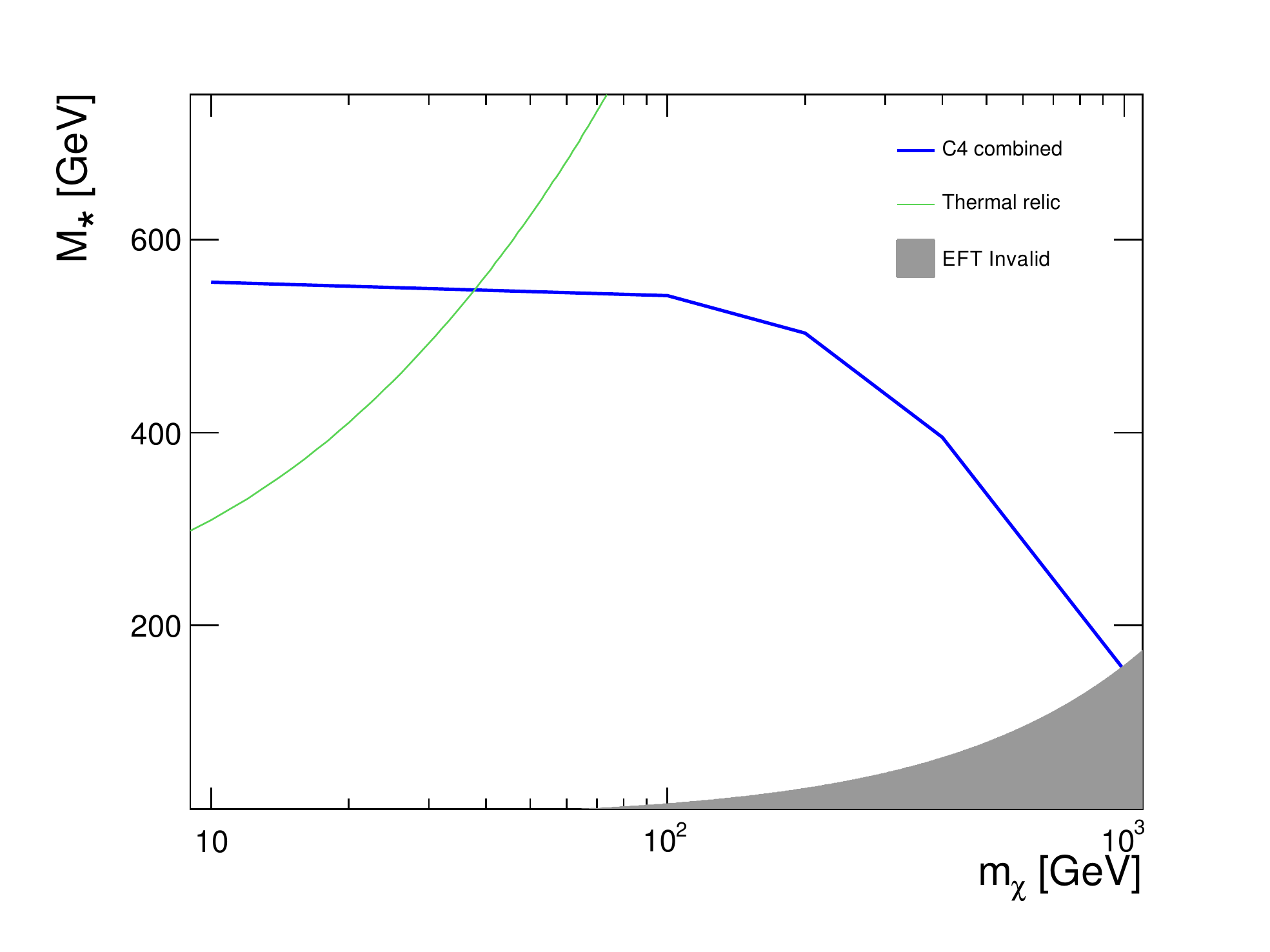}
\includegraphics[width=2in]{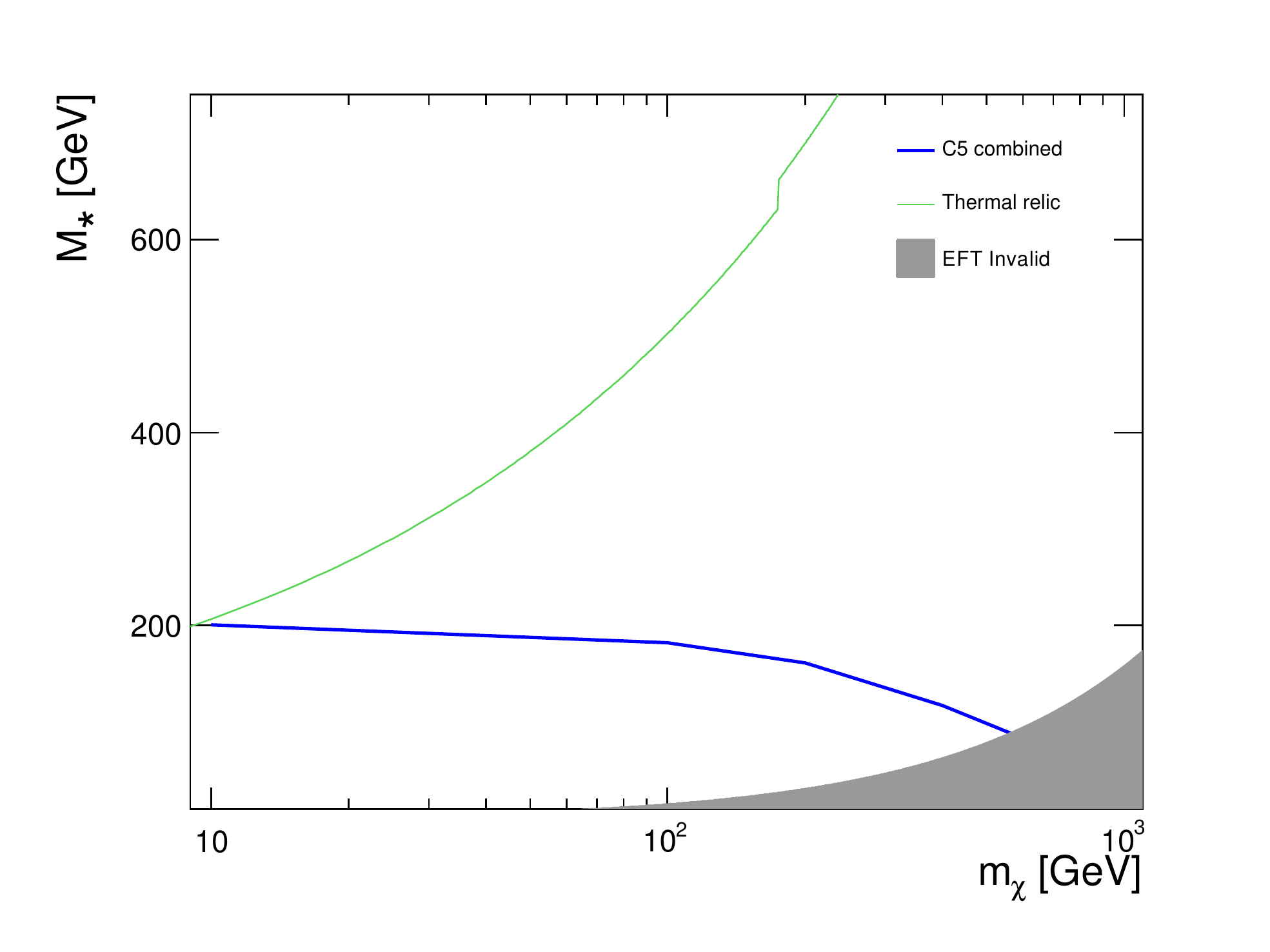}
\includegraphics[width=2in]{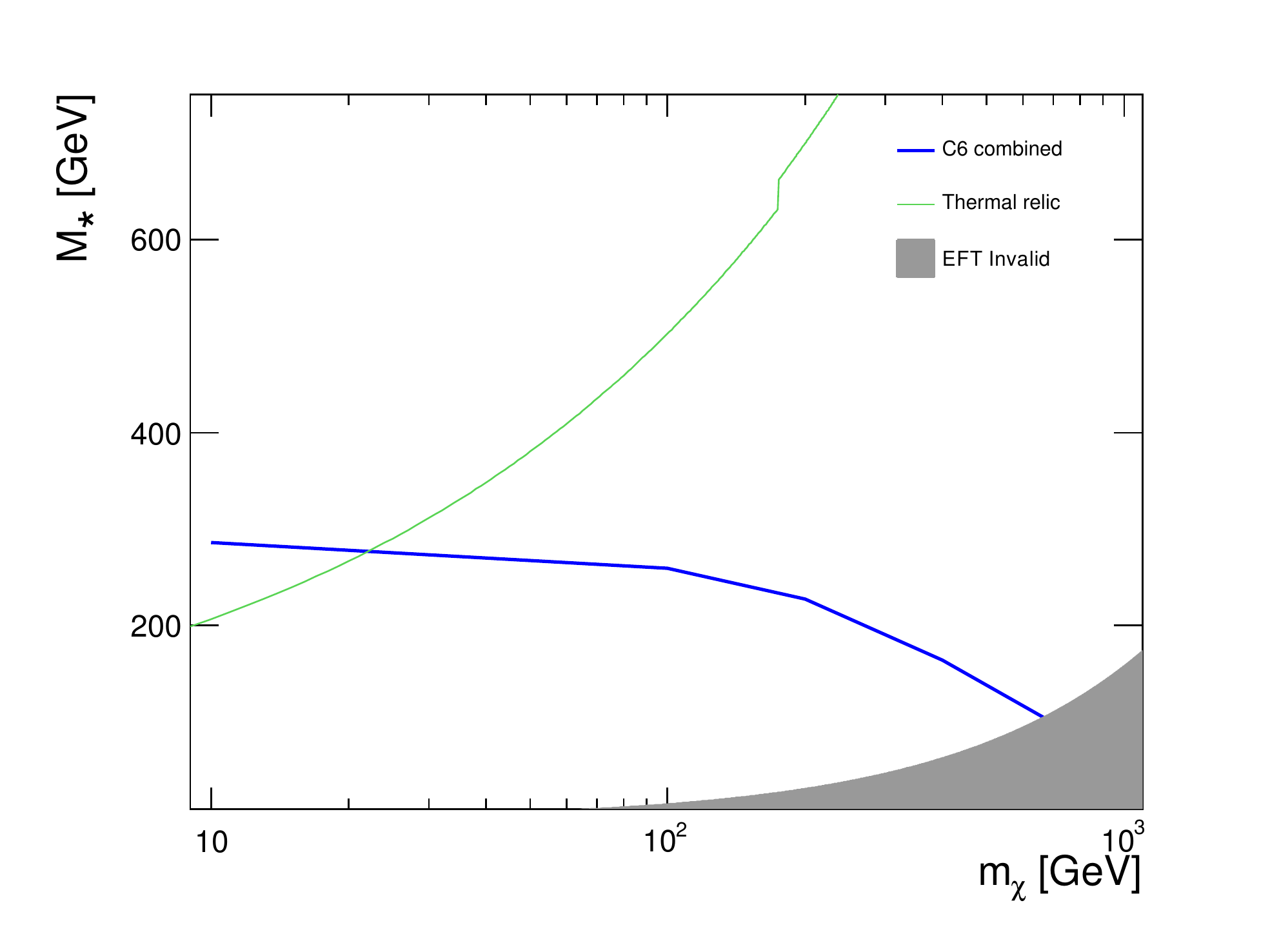}
\caption{ Combined limits on $M_\star$ versus dark matter mass
  $m_\chi$ for operators C4, C5 and C6. The $M_\star$ values at which dark matter particles of a given
  mass would result in the required relic abundance are shown as
  green lines~\cite{Goodman:2010ku}, assuming annihilation in the early universe
  proceeded exclusively via the given operator.}
\label{fig:simple4b}
\end{figure}

In
Figs.~\ref{fig:simple1},~\ref{fig:simple2a},~\ref{fig:simple2b},~\ref{fig:simple3}~\ref{fig:simple4a},
and~\ref{fig:simple4b} we show the combined limits for each operator,
compared to the thermal relic values. Where the limits exceed the
thermal relic values, assuming that dark matter is entirely composed
of thermal relics, the dark matter density of the universe would
contradict measurements and hence cannot couple to quarks or gluons
exclusively via the given operator. This $m_\chi$ region is either
excluded, or else other annihilation channels to leptons must exist,
or finally different operators may interfere negatively thereby
reducing the limits on $M_\star$.

\end{document}